\documentclass[conference]{IEEEtran}
\IEEEoverridecommandlockouts
\usepackage{cite}
\usepackage{amsmath,amssymb,amsfonts}
\usepackage{algorithmic}
\usepackage{graphicx}
\usepackage{textcomp}
\usepackage{xcolor}
\def\BibTeX{{\rm B\kern-.05em{\sc i\kern-.025em b}\kern-.08em
    T\kern-.1667em\lower.7ex\hbox{E}\kern-.125emX}}
\usepackage{mathptmx} 
\usepackage{fancyhdr}
\usepackage[normalem]{ulem}
\usepackage[hyphens]{url}
\usepackage{microtype}
\usepackage{adjustbox}
\usepackage{threeparttable}
\usepackage{textgreek}
\usepackage[bookmarks=true,breaklinks=true,letterpaper=true,colorlinks,linkcolor=blue,citecolor=blue,urlcolor=black]{hyperref}
\usepackage{fancyhdr}
\usepackage[normalem]{ulem}
\usepackage{epsfig}
\usepackage{graphicx}
\usepackage{epstopdf}
\usepackage{setspace}
\usepackage{subfig}
\usepackage[ruled,vlined,linesnumbered]{algorithm2e}
\usepackage{url}
\usepackage{breakurl}
\usepackage{amsmath}
\usepackage{pifont}
\usepackage{fixltx2e}
\usepackage{anyfontsize}
\usepackage{kantlipsum}
\usepackage{multirow}
\usepackage[font=bf,skip=2pt]{caption}
\usepackage{array}
\usepackage{comment}
\usepackage{mwe}
\usepackage{enumitem}
\usepackage{stackengine}
\usepackage{tikz}
\usepackage{kantlipsum}

\newcommand{\mycomment}[1]{}
\newcommand{\ignore}[1]{}

\usepackage{color}
\usepackage{soul}
\usepackage{xspace}

\begin{document}

\title{Amber\textsuperscript{*}: Enabling Precise Full-System Simulation with Detailed Modeling of All SSD Resources
\thanks{\textsuperscript{*}Amber is the project name for SimpleSSD 2.0. All the sources of our simulation framework are available to download from https://simplessd.org.}
\thanks{\textsuperscript{*}\textit{This paper has been accepted at the 51st Annual IEEE/ACM International Symposium on Microarchitecture (MICRO '51), 2018.
This material is presented to ensure timely dissemination of scholarly and technical work.
Please refer and cite the MICRO work of this paper \cite{amber}}}
}

\author{Donghyun Gouk$^{1}$, Miryeong Kwon$^{1}$, Jie Zhang$^{1}$, Sungjoon Koh$^{1}$, Wonil Choi$^{1,2}$,\\ Nam Sung Kim$^{3}$, Mahmut Kandemir$^{2}$ and Myoungsoo Jung$^{1}$\\
{\emph{Yonsei University,}} \\
{\emph{$^{1}$Computer Architecture and Memory Systems Lab, }} \\
{\emph{$^{2}$Pennsylvania State University,}} {\emph{$^{3}$University of Illinois Urbana-Champaign,}}\\
{\small{http://camelab.org}} \\
}

\maketitle

\begin{abstract}
SSDs become a major storage component in modern memory hierarchies, and SSD research demands exploring future simulation-based studies by integrating SSD subsystems into a full-system environment.
However, several challenges exist to model SSDs under a full-system simulations; SSDs are composed upon their own complete system and architecture, which employ all necessary hardware, such as CPUs, DRAM and interconnect network.
Employing the hardware components, SSDs also require to have multiple device controllers, internal caches and software modules that respect a wide spectrum of storage interfaces and protocols. These SSD hardware and software are all necessary to incarnate storage subsystems under full-system environment, which can operate in parallel with the host system.

In this work, we introduce a new SSD simulation framework, SimpleSSD 2.0, namely \emph{Amber}, that models embedded CPU cores, DRAMs, and various flash technologies (within an SSD), and operate under the full system simulation environment by enabling a data transfer emulation. Amber also includes full firmware stack, including DRAM cache logic, flash firmware, such as FTL and HIL, and obey diverse standard protocols by revising the host DMA engines and system buses of a popular full system simulator's all functional and timing CPU models (gem5). The proposed simulator can capture the details of dynamic performance and power of embedded cores, DRAMs, firmware and flash under the executions of various OS systems and hardware platforms. Using Amber, we characterize several system-level challenges by simulating different types of full-systems, such as mobile devices and general-purpose computers, and offer comprehensive analyses by comparing passive storage and active storage architectures.

\end{abstract}


\section{Introduction}
\label{sec:intro}
Non-volatile memories (NVMs) and flash-based storage are widely used in many computing domains, and are changing the role of storage subsystems in modern memory/storage systems. For example, solid state drives (SSDs) have already replaced most conventional spinning disks in handhelds and general computing devices \cite{merry2010solid,dirik2009performance}. Further, servers and high performance computers leverage SSDs as a cache or as a burst buffer in hiding long latencies imposed by the underlying disks \cite{liu2012role,bhimji2016accelerating,bent2012storage}, or placing hot data for satisfying quality of service (QoS) and service level agreement (SLA) constraints \cite{ari2006performance,zhang2010adaptive}.


While most of the time simulation-based studies are necessary to explore a full design space by taking into account different SSD technologies, it is non-trivial to put SSD-based subsystems into a full-system simulation environment. First, even though SSDs are considered as storage or memory subsystems, in contrast to conventional memory technologies, they are in practice incarnated on top of a complete, independent, and complex system that has its own computer architecture and system organization. Second, SSDs consist of not only a storage backend (by having multiple flash packages through internal system buses), but also a computation complex that employs embedded CPU cores, DRAM modules, and memory controllers. As all these hardware components operate in parallel with the host, the simulated evaluations can be easily far from the actual performance. Third, SSDs serve I/O requests and communicate with the host system through different types of storage interfaces,  including serial ATA (SATA) \cite{sata}, universial flash storage \cite{ufs}, NVM Express (NVMe) \cite{nvme}, and open-channel SSD (OCSSD) \cite{ocssd}. Since SSDs are subservient to the host system and OS decisions, SSD-based subsystem or full-system simulations without considering these storage interfaces and software stacks can result in ill-tuned and overly-simplified evaluations. Lastly, SSDs also employ different firmware modules optimized to reduce the performance disparity between the host interface and the internal storage complex, which also have a great impact on user-level system behavior, based on different workload executions on full-system environment.


Unfortunately, most SSD simulation infrastructures available today have no full software stack simulation through emulation of data transfers and they also lack SSD internal hardware and/or software resource models, which makes them insufficient to be put into a full-system environment.
For example, \cite{ssd-extension}, which is widely used for SSD simulations in both academia and industry, only captures the functionality of a specific flash firmware, called flash translation layer (FTL) without modeling any hardware resource in an SSD. Similarly, most recent simulators \cite{simplessd, simplessd-x, mqsim} have no model for computation complex and therefore, no detailed timing simulation for firmware execution can be carried out. Further, to the best of our knowledge, there exists no simulation model that can be integrated into a full-system environment by implementing all actual storage interfaces and data transfer emulations.
For example, \cite{mqsim} only mimics a pointer-based multi-queue protocol (in storage) without modeling system buses and data movements. As a result, it cannot be integrated into a full system setting that requires the emulating a real OS and all relevant software/hardware components.
Lastly, none of the existing SSD simulators can be attached to different storage interfaces by respecting \textit{both} functional and timing CPU modeled full-system infrastructures.
For example, \cite{simplessd} and \cite{simplessd-x} enable a full-system simulation, but they only work on a functional CPU, which overly simplifies the host memory subsystem and CPU execution timings.



Motivated by the lack of full system simulation tools that can accommodate SSDs with all functional and timing parameters as well as firmware, we introduce a new SSD simulation framework, SimpleSSD 2.0, namely \emph{Amber}, which accommodates all SSD resources in a full system environment and emulates all software stacks employed by both functional and timing CPU models. The proposed simulation framework modifies host-side system buses, and implements device-side controllers and a DMA engine to emulate data transfers by considering a wide spectrum of storage interfaces and protocols, such as SATA, UFS, NVMe, and OCSSD. In addition, Amber implements a diverse set of firmware modules on top of detailed SSDs computation/storage complex models, and applies different flash optimizations such as parallelism-aware readahead \cite{chen2009understanding} and partial data update schemes \cite{abraham2016partial}, which allows it to easily mimic the performance characteristics of real systems. While most existing SSD simulators evaluate the performance of an underlying storage complex by replaying block-level traces, we evaluate/validate Amber by actually executing different microbenchmarks as well as user-level applications on real OS-enabled systems.


To the best of our knowledge, \emph{Amber} is the first SSD simulation framework, that incorporates \emph{both} computation and storage complexes within an SSD and covers a large and diverse set of storage interface protocols and data transfer emulation, while considering a full-system storage stack. The main contributions of this work can be summarized as follows:

\noindent $\bullet$ \emph{Hardware and software co-simulation for storage.}
For SSD's computation complex, we model and integrate embedded CPU cores, DRAM modules/controller and system buses, which can capture the detailed latencies and throughput of the execution of flash firmware components based on ARMv8 ISA. Amber's storage complex implements reconfigurable interconnection networks that can simulate a wide spectrum of flash technologies by considering detailed flash transaction timing models. We integrate a full firmware stack on top of the computation and storage complexes, thereby capturing diverse SSD functionalities, such as I/O caching, garbage collection, wear-leveling, and address translation. This co-simulation framework offers realistic storage latency and throughput values by incorporating all {\em SSD} resources into a full-system. Amber can also be used to explore various dynamics of an SSD, which are critical for monitoring power/energy usages of different storage components under real application execution scenarios.

\noindent $\bullet$ \emph{Enabling SSDs in diverse full-system domains.}
Real systems can attach an SSD over different locations based on their platform and user demands. We enable both I/O controller hub's hardware-driven storage (e.g., SATA and UFS) and memory controller hub's software-driven storage (e.g., NVMe and OCSSD) by implementing all their mandatory commands and data transfer mechanisms. To this end, we integrate our hardware/software co-simulation storage models into gem5 \cite{gem5}, and revise gem5 for tight integration with data transfer emulation. Amber modifies gem5's system bar and models a host-side DMA engine that moves data between the OS system memory and the underlying storage. This work also includes a set of protocol-specific implementations such as host/device controllers, system memory references over a pointer list, and a queue arbitration logic. Amber works with all of gem5's CPU types such as timing and functional CPU models for both in-order and out-of-order executions under different OS versions.


\ignore{
\noindent $\bullet$ \emph{Firmware optimizations and hardware tuning.}
One of the challenges behind SSD simulations is to catch up the realistic performance by considering system-level I/O design parameters. We observe that state-of-the-art simulators exhibit similar performance compared to real device under a limited execution environment, and their performance behaviors are far from actual systems in cases where one considers different block sizes and I/O depths. This is because most real SSD devices have several optimizations applied into internal DRAM caches and FTL, and the computation complex, and they contribute performances in diverse perspectives. We tailor data caching and FTL software modules being aware of internal resources and I/O parallelism of the underlying storage complex. We validate that the performance trends of Amber are very similar to real systems by executing actual applications that performs the tests with various I/O depths and block sizes at user level.
}
\noindent $\bullet$ \emph{Holistic analysis for different storage subsystem designs.} Since Amber can emulate data transfers and execute a full storage stack, including host's system software and device's firmware, We implement a diverse set of SSD subsystems in Amber, and analyze their system-level challenges. Specifically, we characterize i) the performance impact of employing different operating systems, ii) mobile system challenges regarding UFS/NVMe, and iii) passive and active storage architectures. We observe that OS-level ill-tuned I/O scheduling and queue management schemes can significantly degrade the overall system performance, and that user applications in mobile computing should be revised by considering high performance SSD designs. Specifically, even  though NVMe (attached ARM core) exhibits much better performance than UFS, some applications cannot take advantage of NVMe due to their default I/O operation mode. Our evaluations also reveal that the CPU and memory utilization can be problematic issues that a passive storage architecture (employing host-side FTL over OCSSD) needs to address in the future. In particular, while the active storage (having NVMe controller and full firmware stack in an SSD) consumes only 7\% of the host-side CPU, the passive storage consumes most of host-side resources.






\section{Background}
\label{sec:background}
\begin{figure*}
	\centering
	\begin{minipage}[b]{.57\linewidth}
		\subfloat[Overview.\label{fig:bck_overview}]{\includegraphics[origin=c,width=0.33\linewidth]{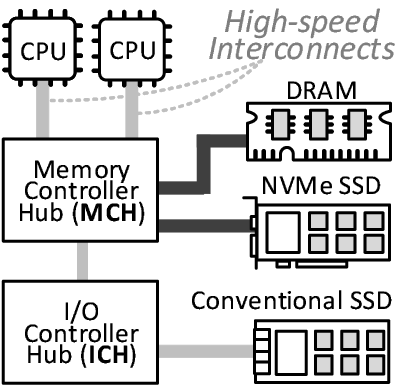}}
		\subfloat[H-type.\label{fig:bck_htype}]{\includegraphics[origin=c,width=0.352\linewidth]{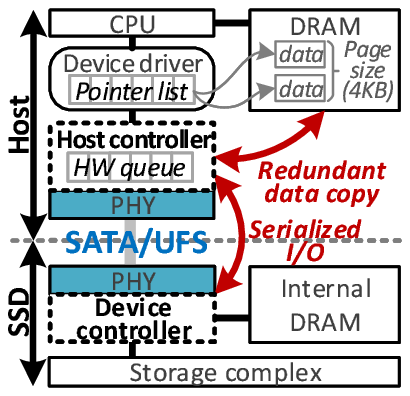}}
		\subfloat[S-type.\label{fig:bck_stype}]{\includegraphics[origin=c,width=0.308\linewidth]{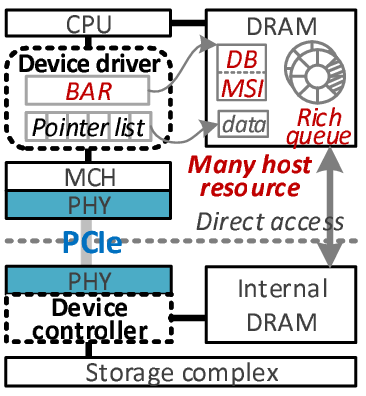}}
		\caption{System architecture\label{fig:bck_arch}.}
	\end{minipage}
	\begin{minipage}[b]{.42\linewidth}
		\stackunder[8pt]{\includegraphics[width=1\linewidth]{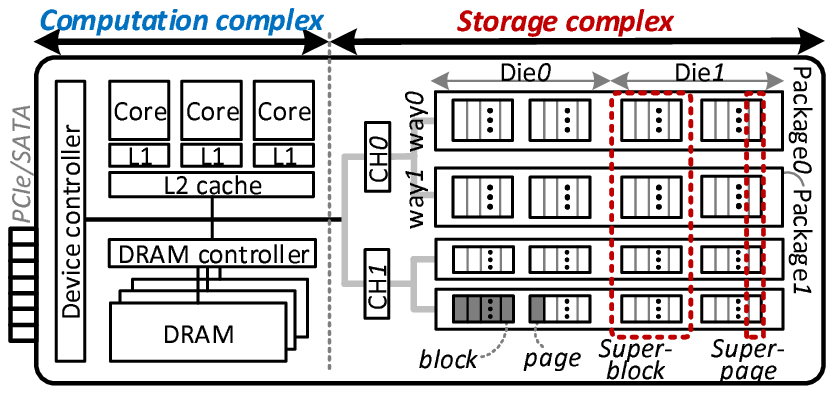}}{ }
		\caption{\label{fig:bck_ssdhw}SSD internal architecture.}
	\end{minipage}
	\vspace{-15pt}
\end{figure*}

\subsection{System Architecture}

\noindent \textbf{Overview.} SSDs can be attached to modern systems by memory controller hub (MCH) and/or I/O controller hub (ICH) via PCI Express (PCIe) or serial I/O ports, respectively. As shown in Figure \ref{fig:bck_overview}, MCH is directly connected to CPU via a front-side bus, and usually manages high-speed I/O components, such as DRAM, GPGPU and NVMe SSD devices. In contrast, ICH is paired with MCH, but handles relatively slow devices, such as spinning disks and conventional SSDs. Due to the different purposes of I/O connection mechanisms of MCH and ICH, in practice, SSDs and their interfaces can classified into two types of storage subsystems: i) \emph{hardware-driven storage} (h-type) and ii) \emph{software-driven storage} (s-type). h-type storage includes serial ATA (SATA) and universal flash storage (UFS) SSDs, whereas NVM Express SSDs (NVMe) and open-channel SSDs (OCSSD) operate as s-type storage. Overall, h-type storage is more efficient than s-type storage in terms of both I/O management efficiency and host-side resource utilization, but its latency and bandwidth are somewhat limited, preventing it from taking full advantage of the underlying flash media, due to frequent hardware interventions and queue/interrupt management complexities.


\noindent \textbf{H-type storage.} As shown in Figure \ref{fig:bck_htype}, this type of storage employs both a ``host'' controller and a ``device'' controller for data communication over the SATA or UFS interface protocols. The host system's CPU and device drivers can issue an I/O request to the underlying storage and manage interrupt service routine (ISR) only through the host controller. This host controller resides in ICH, and unpacks/packs multiple commands and payload data by using its own buffer. The host controller sends/retrieves data to/from the device controller that exists within an SSD over its physical layer (PHY) on behalf of the host's OS drivers. There are two main architectural challenges behind modeling h-type storage. First, the data transfers between the host and underlying storage require multiple data copies since CPU only configures a set of registers to issue an I/O request. Specifically, for data communication, the drivers (running on CPU) construct a \emph{pointer list} (each entry indicating the target system memory (DRAM) page), rather than transferring contents between the host and the storage. Therefore, the host controller should traverse all memory pointers and copy each system memory page to its own buffer. During this phase, the host controller issues a request by communicating with the target device controller through different PHYs. Second, since only the host controller manages data movements, the number I/O requests controlled by its hardware queue and interrupt mechanism is comparably small and limited. In addition, all the I/O queue management and interrupt handling of the host should be serialized with the h-type storage architecture.
This single I/O path is often considered as a major performance bottleneck in many computing domains \cite{walker2012comparison, kim2015improving}.

\noindent \textbf{S-type storage.} This type of storage has no host controller, and each component of the host is managed by software. Specifically, s-type storage's device controller exposes specific parts of the internal DRAM to the host's designated system memory regions via PCIe, which is referred to as \emph{base line address} (BAR) \cite{bar}. The software modules in the host-side storage stack, such as host NVMe or OCSSD drivers, can directly configure such memory-mapped BAR spaces. Since memory-mapped  spaces are in parallel visible to the device controller, SSD can also directly pull or push data over the host-side system memory pages. In addition, interrupts in the s-type storage are managed by a request packet, referred to as \emph{message-signaled interrupt} (MSI/MSI-X) \cite{msi}. Once the device controller completes an I/O service, it directly writes the interrupt packet into another memory-mapped region (on the host DRAM), called the MSI/MSI-X vector. While the OS driver and device controller can manage I/O request submissions and completions through BARs and MSI vector, it is non-trivial to make their I/O queues consistent and coherent between the host and storage. To address this challenge, s-type storage supports per-queue head and tail pointers, which can be written into doorbell registers by OS drivers but exposed by the device controller. With the head and tail pointer mechanisms in place, the host device driver and device controller can synchronize enqueue and dequeue status for each I/O submission and completion. This data communication method allows s-type storage to employ a large number of queues and queue entries (referred to as \emph{rich queue}), and makes the host eliminate the I/O queue as well as the ISR serialization issues. However, since the software modules are now involved in data transfers, s-type storage requires more host-side resources than h-type storage \cite{jung2016exploring, jung2013challenges, zhang2014power}.

\subsection{SSD Internals}

\begin{figure*}
\centering
	\begin{minipage}[b]{.75\linewidth}
\subfloat[Sequential read.]{\label{fig:sr_bw}\rotatebox{0}{\includegraphics[width=0.24\linewidth]{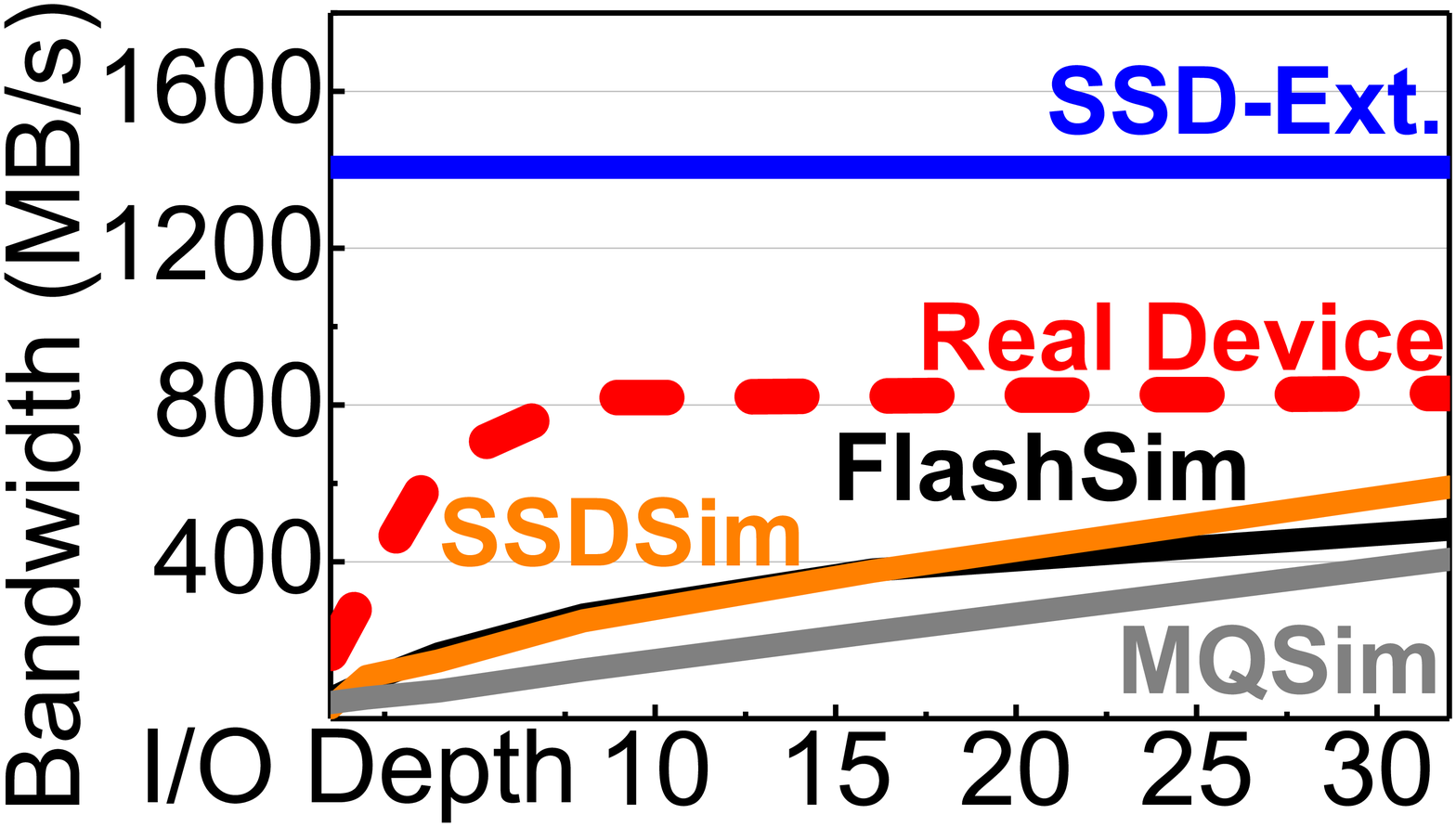}\hspace{5pt}}}
\subfloat[Random read.]{\label{fig:rr_bw}\rotatebox{0}{\includegraphics[width=0.24\linewidth]{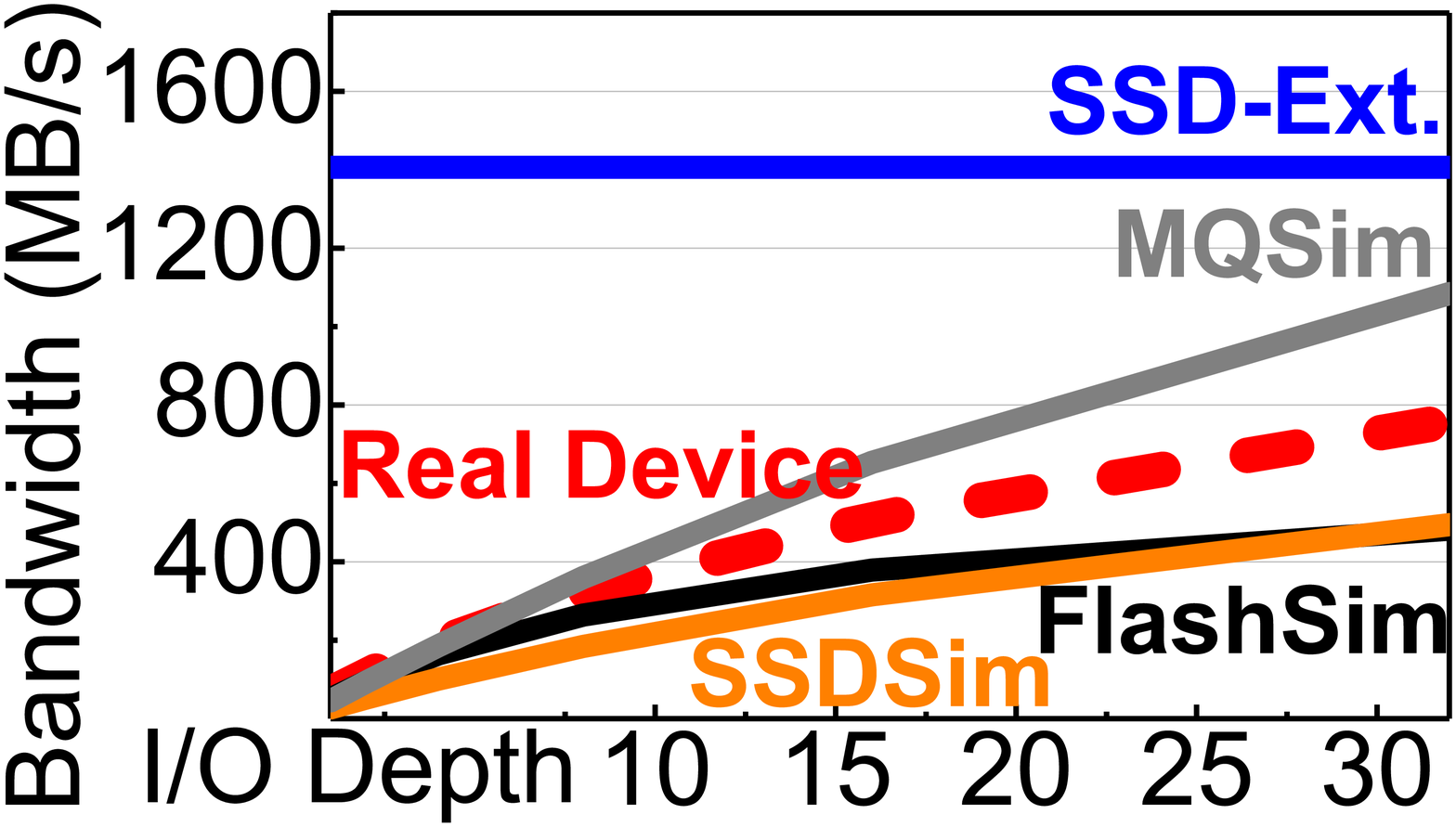}\hspace{5pt}}}
\subfloat[Sequential write.]{\label{fig:sw_bw}\rotatebox{0}{\includegraphics[width=0.24\linewidth]{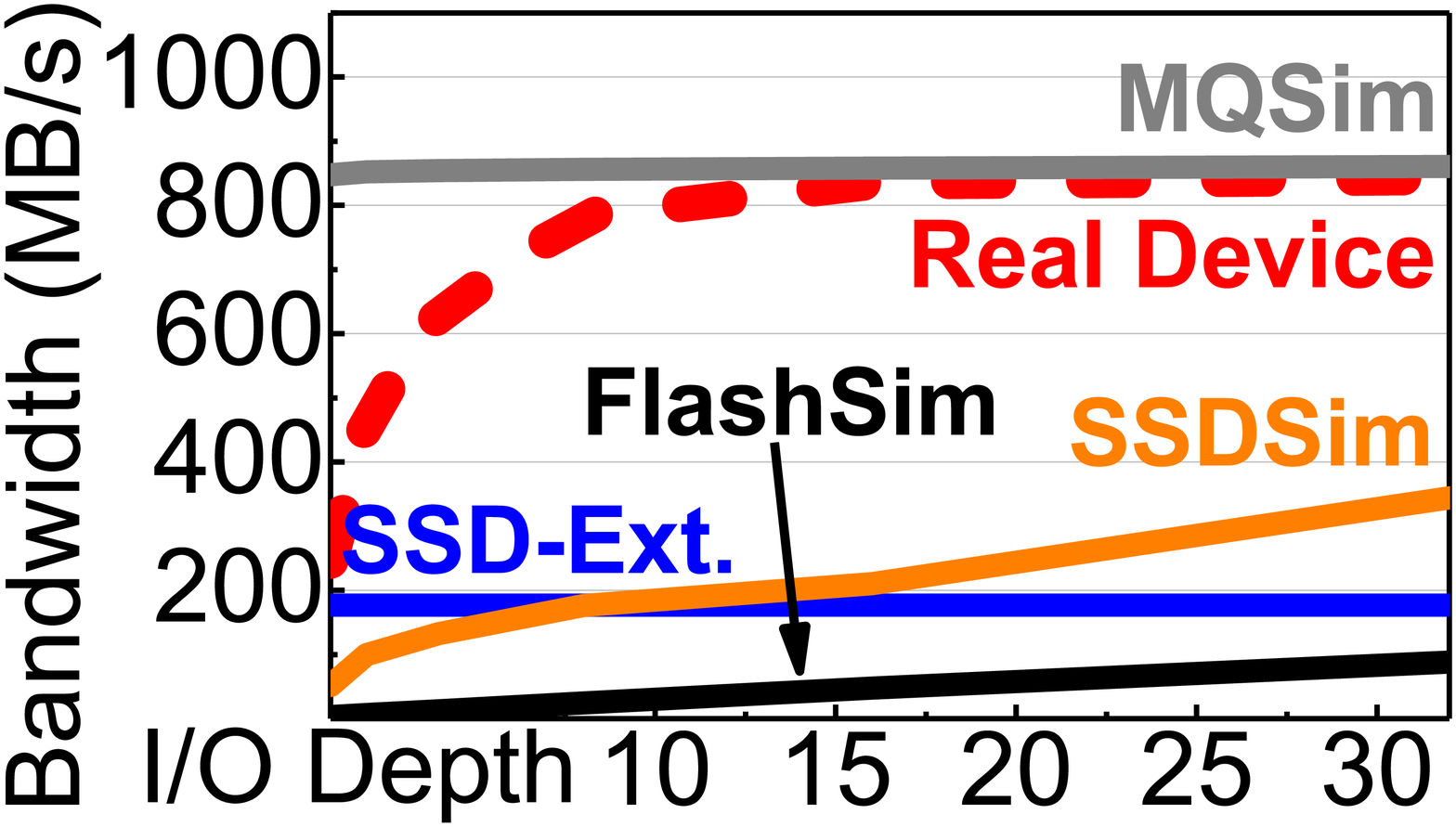}\hspace{5pt}}}
\subfloat[Random write.]{\label{fig:rw_bw}\rotatebox{0}{\includegraphics[width=0.24\linewidth]{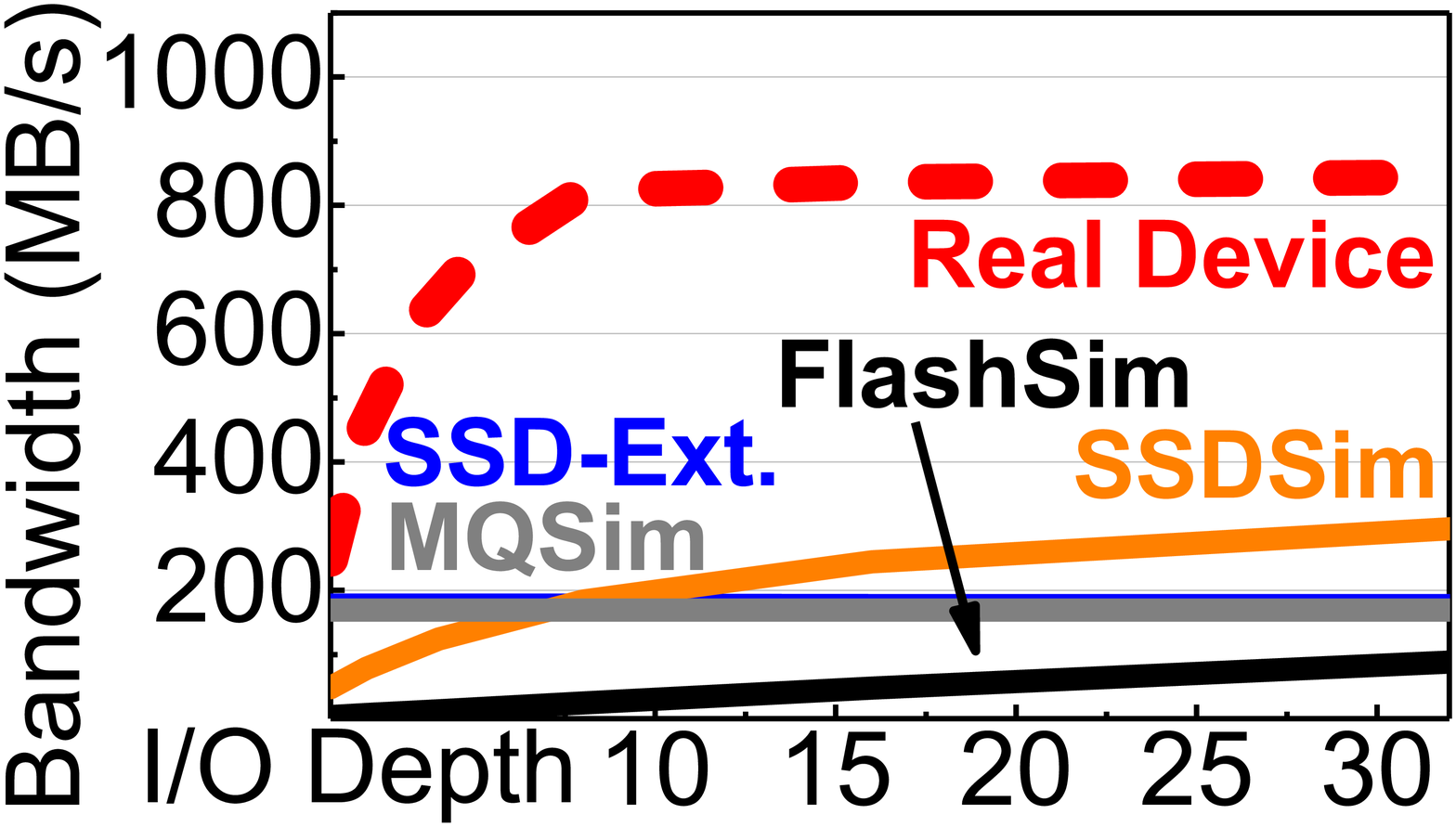}}}
\caption{Bandwidth comparison between a real device and existing SSD simulators.}
\label{fig:bw_comp}
\vspace{8pt}
\subfloat[Sequential read.]{\label{fig:sr_lat}\rotatebox{0}{\includegraphics[width=0.24\linewidth]{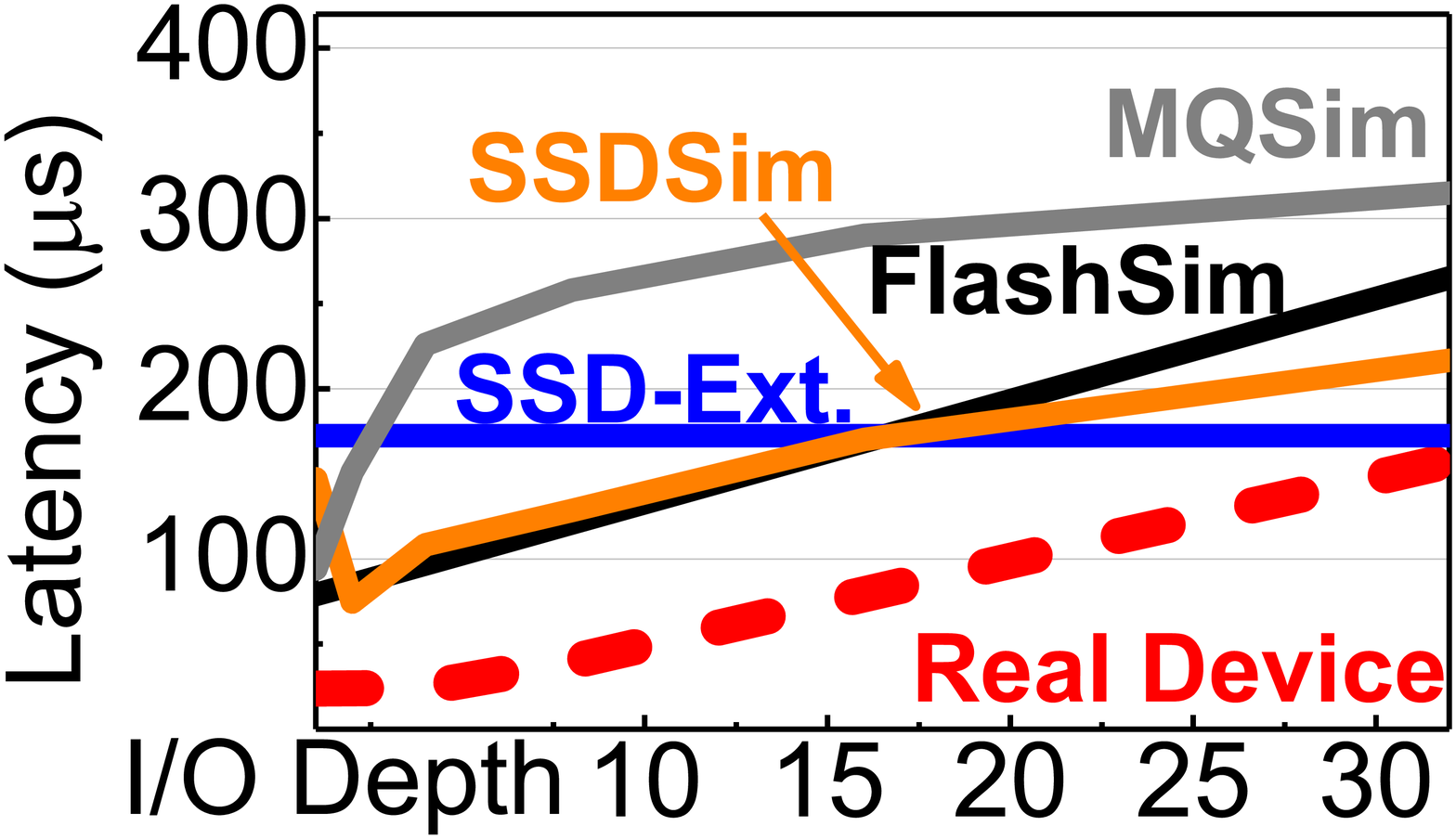}\hspace{5pt}}}
\subfloat[Random read.]{\label{fig:rr_lat}\rotatebox{0}{\includegraphics[width=0.24\linewidth]{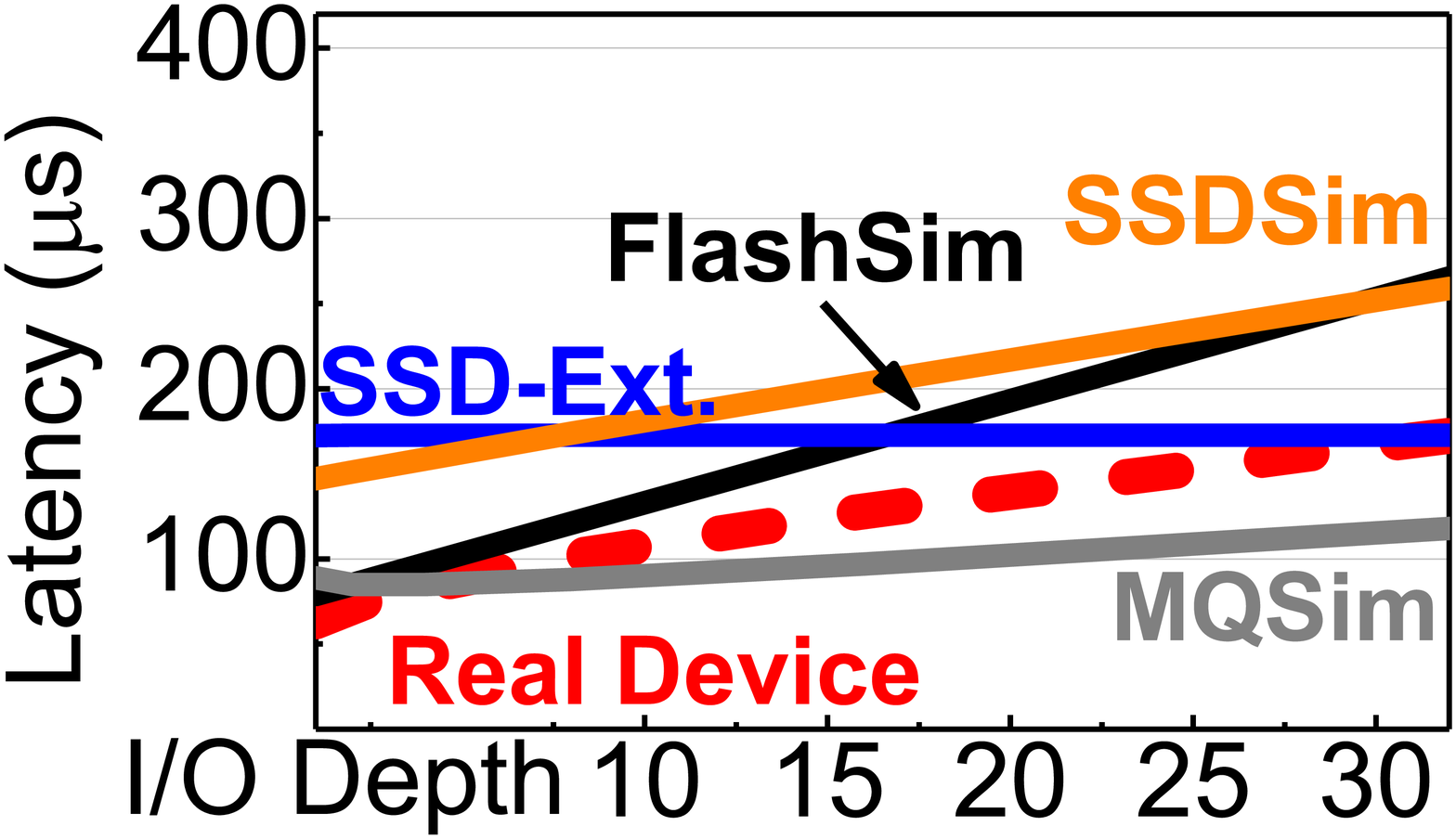}\hspace{5pt}}}
\subfloat[Sequential write.]{\label{fig:sw_lat}\rotatebox{0}{\includegraphics[width=0.24\linewidth]{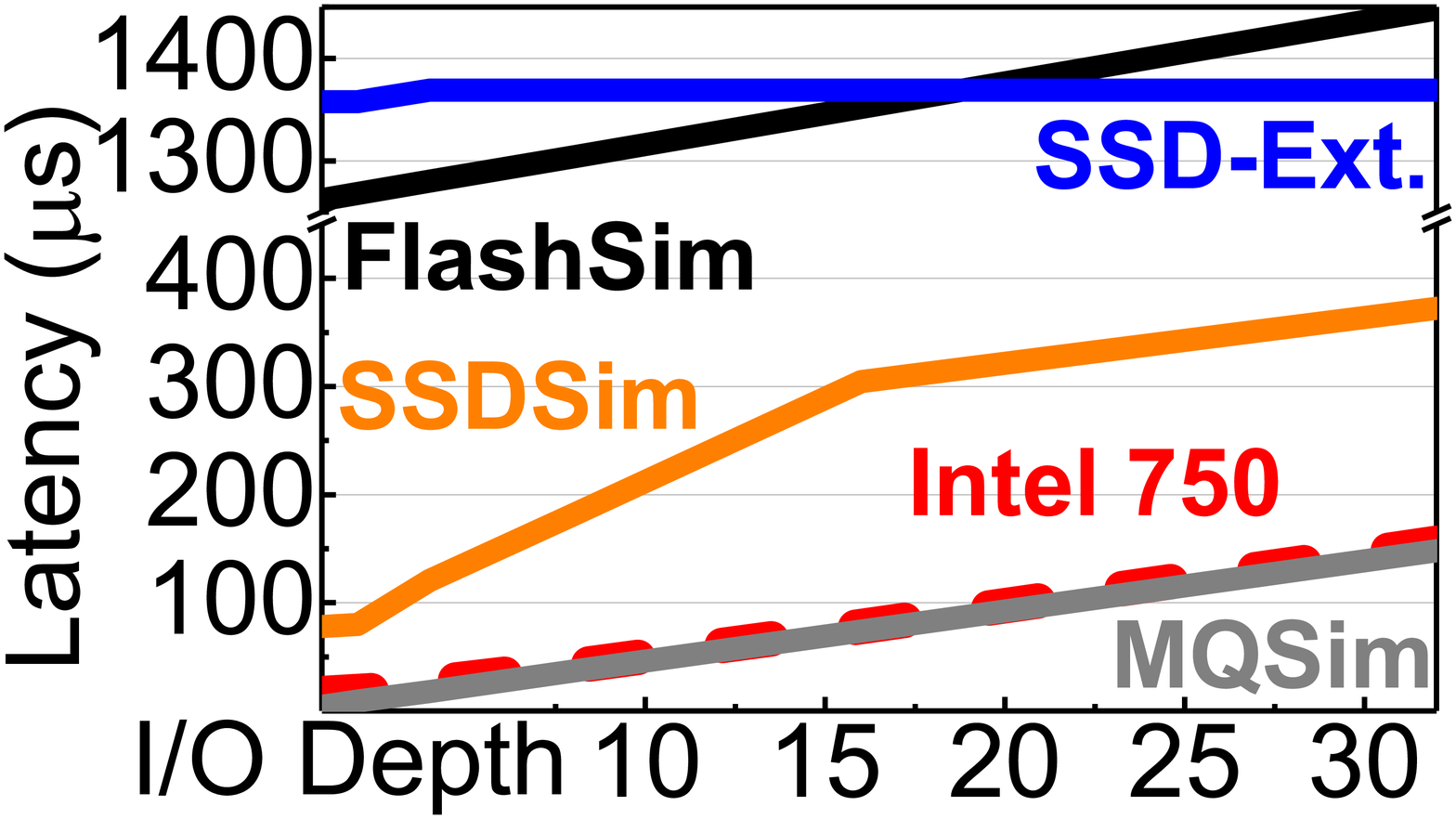}\hspace{5pt}}}
\subfloat[Random write.]{\label{fig:rw_lat}\rotatebox{0}{\includegraphics[width=0.24\linewidth]{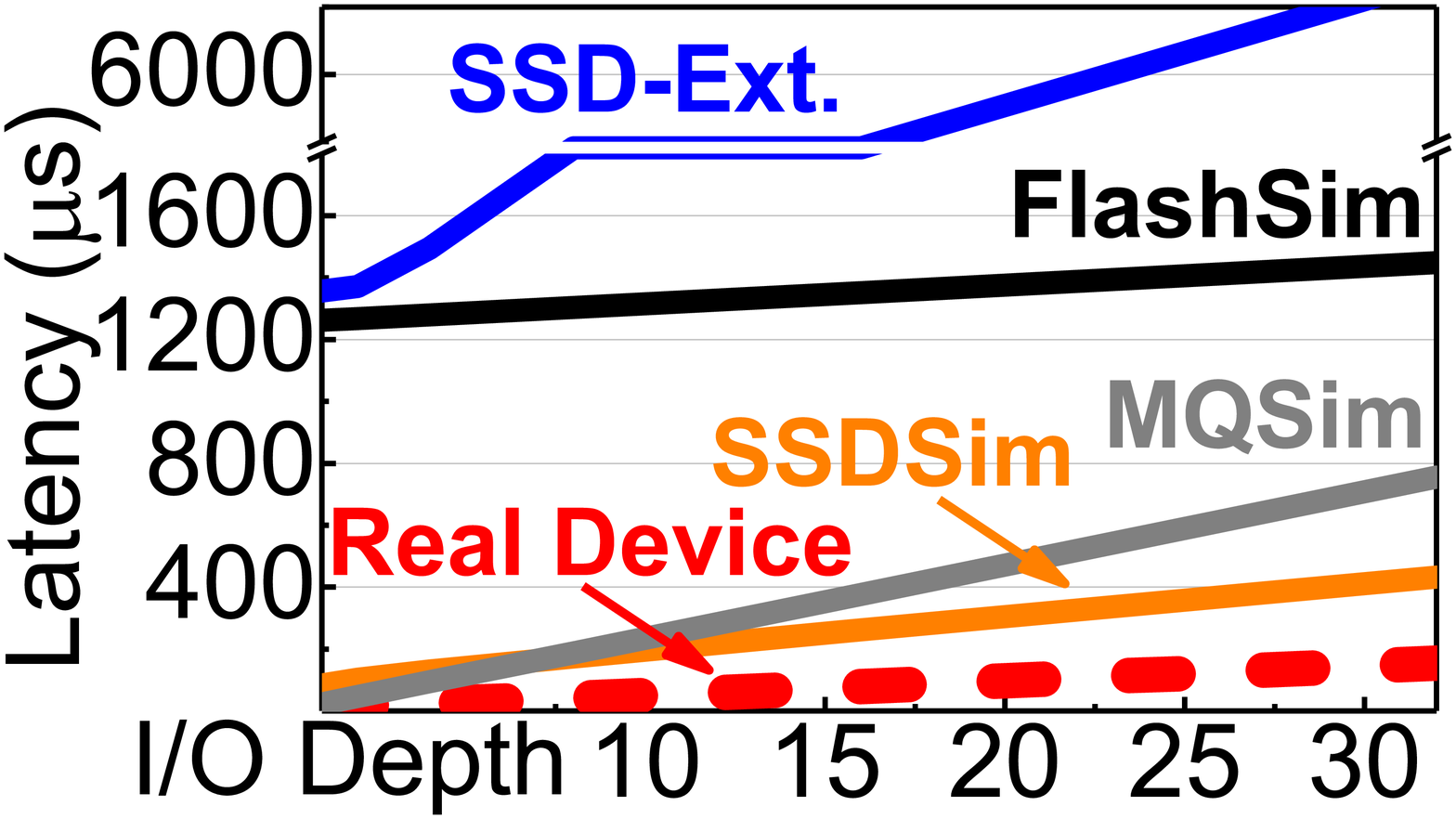}}}
\caption{Latency comparison between a real device and existing SSD simulators.}
\label{fig:lat_comp}
	\end{minipage}
	\begin{minipage}[b]{.23\linewidth}
\centering
\fontsize{8pt}{8.5pt}
\selectfont
\begin{tabular}{|c|c|c|}
\hline
\multicolumn{3}{|c|}{\textbf{NAND Flash timing (\textmu s)}} \\ \hline
{$t\textsubscript{PROG}$} & 820.62 & 2250  \\ \hline
{$t\textsubscript{R}$ } & 59.975 & 104.956 \\ \hline
{$t\textsubscript{ERASE}$ } & \multicolumn{2}{c|}{3000}  \\ \hline
\multicolumn{3}{|c|}{\textbf{Storage back-end}}                                               \\ \hline
Channel & Package & Die \\ \hline
12 & 5  & 1 \\ \hline
Plane & Block    & Page   \\ \hline
 2  & 512   & 512  \\ \hline
\multicolumn{3}{|c|}{\textbf{Internal DRAM}} \\ \hline
Size        & Channel    & Rank  \\ \hline
1GB         & 1          & 1     \\ \hline
Bank & Chip & Bus width \\ \hline
8    & 4    & 8         \\ \hline
\end{tabular}
\vspace{2pt}
\captionof{table}{Hardware configuration of real device.\label{tab:hwconfig}}
	\end{minipage}
	\vspace{-3pt}
\end{figure*}

\noindent \textbf{Hardware architecture.} Since the performance of flash media is much lower than the bandwidth capacity of host storage interfaces, most modern SSDs are designed based on a multi-channel and multi-way architecture to improve internal parallelism, with the goal of reducing the bandwidth gap between the host interface and flash \cite{chen2011essential,jung2012evaluation}. As shown in Figure \ref{fig:bck_ssdhw}, multiple flash packages, each containing multiple dies, are connected to the interconnection buses, referred to as \emph{channel}. The set of packages across different channels can simultaneously operate, and in practice, flash firmware spreads a host request over multiple dies that have the same offset address, but exist across different channels. Each set of flash dies (or packages) is called a \emph{way}. A group of multiple physical pages and blocks that span over all internal channels or ways is referred to as \emph{super-page} or \emph{super-block}, respectively. In this multi-channel and multi-way architecture, the \textit{storage complex} (putting all flash media into the interconnection network) is also connected to a \textit{computation complex} that consists of embedded CPU cores, internal DRAM, and device controller(s). Multiple cores are allocated to flash firmware that controls all I/O services and address translations within an SSD. Since the operating frequency domains between the host and storage are completely different, all data coming from the underlying flash (e.g., reads) or the host-side driver or controller (e.g., writes) should be first buffered in the internal DRAM and should then be moved to the target device.
Note that the flash media of the storage complex allows no overwrite, due to the \emph{erase-before-write} characteristic/requirement. Specifically, at the flash-level, writes/reads are served per page, whereas erases should be processed per block. In practice, a flash block contains 128 $\sim$ 512 pages, and the latency of an erase operation is 50 times longer than that of a page write. In addition, the order of writes in a block should be \textit{in-order} to avoid page-to-page interference/disturbance for the multi-level cell (MLC) and triple-level cell (TLC) flash technologies.

\noindent \textbf{Software organization.} Due to the  aforementioned erase-before-write characteristic, all existing flash-based storage today require employing a firmware that makes it compatible with the conventional block storage and hides the complexity of flash management. \emph{Flash Translation Layer} (FTL) is a key component of the flash firmware that prepares a number of blocks, which are erased in advance (called \emph{reserved block}). FTL writes data into a reserved block by mapping the input address (\emph{logical block address}, LBA) to \emph{physical page address/number} (PPA/PPN). In cases where there is no reserved block at runtime, FTL migrates valid physical pages from old block(s) to a new block, erases the old block, and remaps the addresses of the migrated pages, thereby securing available block(s) to forward incoming write requests. This process is called \emph{garbage collection} (GC). Since the number of erases per block is limited due to flash wear-out issues \cite{pan2011exploiting,slcmlc}, when FTL performs a GC task, it tries to erase flash blocks in an evenly distributed manner, referred to as \emph{wear-leveling}. In addition, flash firmware also implements request parsing, protocol management, I/O scheduling, and data caching. Specifically, the device controller that exists for both h-type storage and s-type storage, manages the data communication based on the interface/protocol defined by the host. Then, host interface layer (HIL) transfers data and performs I/O scheduling atop other flash firmware modules. Since the data is buffered in the internal DRAM in an SSD, the flash firmware can also cache the data to leverage DRAM performance \cite{meza2015large,shim2010adaptive}. This in turn can help us hide the long latency imposed the underlying flash media.


\section{Enabling Hardware/Software Co-Simulation}
\label{sec:highlevelview}

\subsection{Challenges in Capturing Realistic SSD Performance}

\begin{figure*}
\captionsetup[subfigure]{captionskip=-2pt}
	\subfloat[Internal architecture.]{\label{fig:high_arch}\rotatebox{0}{\includegraphics[width=0.66\linewidth]{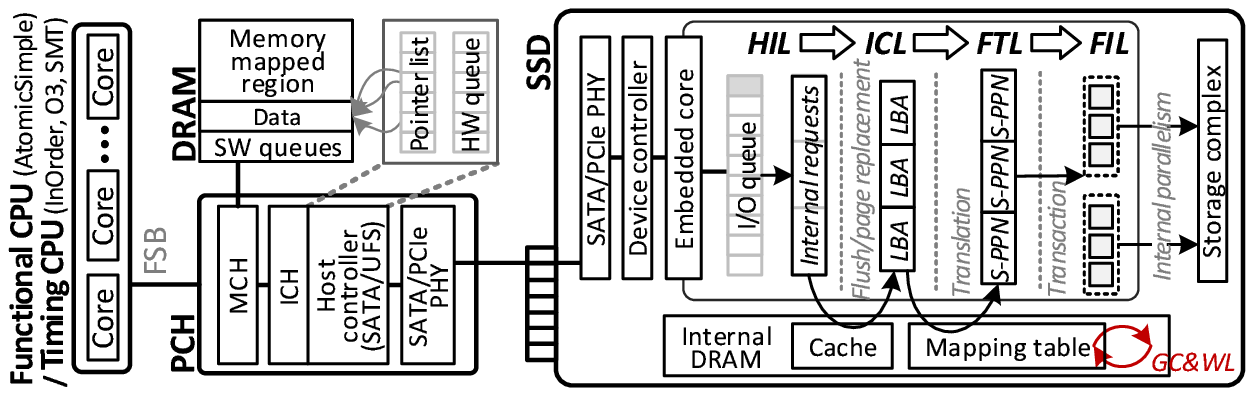}}}
	\hspace{0.5pt}
	\subfloat[Changes for gem5 system barbus and DMA.]{\label{fig:high_full}\rotatebox{0}{\includegraphics[width=0.33\linewidth]{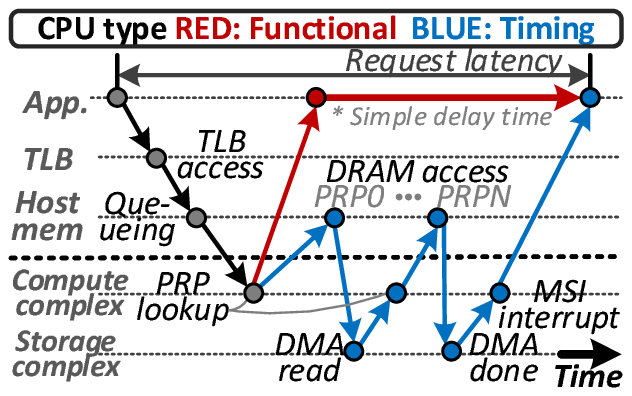}}}
	\caption{High-level view of \texttt{Amber}\label{fig:high_amber}.}
	\vspace{-3pt}
\end{figure*}

Figures \ref{fig:bw_comp} and \ref{fig:lat_comp} compare SSD bandwidth and latency of a real device (Intel 750) and existing SSD simulators, namely, MQSim \cite{mqsim}, SSDSim \cite{ssdsim}, SSD Extension for DiskSim (SSD-Extension) \cite{ssd-extension}, and FlashSim \cite{flashsim}. To perform this analysis, we disassemble an Intel 750 device and reverse-engineer the number of channels, ways, flash dies, and DRAM sizes. We then check the part number of flash and also extract the low-level flash latencies for the read, write and erase operations from datasheets. We then configure all simulators' device parameters by referring the datailed hardware configuration information given in Table \ref{tab:hwconfig}. We extract 4KB-sized block traces from the flexible I/O tester synthetic benchmark (FIO \cite{fio}) and replay the traces on each simulator mentioned above, with I/O depths varying from 1 to 32. Since none of these existing simulators is suitable to execute FIO at user level by having full storage stack over full-system environment, \emph{replaying or generating traces} within their simulation framework is the only way to evaluate their performance with the same device configuration.

\noindent \textbf{Bandwidth trend.}
One can observe from Figure \ref{fig:bw_comp} that, all existing simulators' bandwidths are \textit{far} from the performance of the real device. Specifically, Intel 750 keeps utilizing the bandwidth, and its performance saturates with 8 $\sim$ 16 queue entries (except for random reads). As the number of queue entries increases, most existing simulators exhibit performance curves that are completely different from that of the real device. Specifically, the existing simulators' bandwidth i) linearly increases (MQSim and SSDSim), ii) exhibits just a constant trend (SSD-Extension and FlashSim), or iii) shows a curve but does not saturate (SSDSim). In contrast, the bandwidth trend of the real device is sublinear as the number of queue entries increases. In addition to these trend differences, all the simulators tested exhibit significant performance disparity between their models and the real device. With 16 queue entries (i.e., steady-state), the error in MQSim's reads and writes ranges between 3\% and 80\%, compared to Intel 750. Other simulators' error ranges are even more serious. More specifically, the resulting errors with  SSDSim, SSD-Extension, and FlashSim for reads and writes are as high as 54\%, 176\%, and 54\%, and 75\%, 79\%, and 94\%, respectively.

\noindent \textbf{Latency trend.} As shown in Figure \ref{fig:lat_comp}, the latency bahavior is different from the aforementioned bandwidth trends. As the I/O depth is increased, the latency exhibited by the real device gets worse (due to the queueing delays). But, for {\em all} random and sequential I/O patterns, its latency is lower than 175 us. While the existing simulators also show an increased latency as the number of queue entries increases, their latencies range between 4 us and 6285 us, and exhibit different performance curves, compared to the real device. The existing simulators with varying I/O depths exhibit i) sublinear-like latency trend (MQSim, SSDSim and SSD-Extension), ii) constant latency trend (SSD-Extension), and iii) linear trend curved by unrealistic gradients (FlashSim). For example, for sequential writes, MQSim offers a latency behavior similar to that of the real device, and its latency on sequential reads exhibits a sublinear-like behavior when increasing the I/O depth. SSD-Extension shows a sublinear-like latency curve for random writes, but it does not have any latency variance when modulating the number of queue entries under the execution of all other I/O workloads. The latency errors incurred by MQSim, SSDSim, SSD-Extension and FlashSim for reads and writes are as high as 816\%, 518\%, 627\%, and 293\%, and 388\%, 495\%, 8492\%, and 7887\%, respectively.


We believe that the main reasons why most of the existing simulators exhibit significantly different performance trends and values from the real device, with varying I/O depths, are: i) lack of a computation complex, ii)  incomplete firmware stack, iii) omitted storage interface and protocol management, and iv) absence of a host-side I/O initiator (e.g., host driver or controller). In contrast, Amber can simulate both computation and storage complexes, and execute FIO from user-level, employing all hardware and software components on a full-system environment. Overall, Amber exhibits performance trends that are very similar to those of the real device when varying the I/O depth, and further, its user-level latency and bandwidth are different from those observed with the real device by 20\% and 14\%, on average, respectively. A more detailed analysis will be given later in Section \ref{sec:evaluation}.


\subsection{High-Level View of Amber}

\noindent \textbf{Internal architecture.}
Figure \ref{fig:high_arch} depicts the internal architecture that Amber models for an SSD.
Amber models embedded CPU cores based on ARMv8 instruction set architecture (ISA) \cite{armv8}; we decompose the instructions of each module's functions in a fine-granular manner, and allocate each firmware component to different CPU core. The function call procedures and control flow of SSD firmware stack can vary based on the workloads and the OS decisions made by the software emulation of higher full-system environment, Amber keeps track of all dynamic call procedures for each run, and measures instruction executions by considering arithmetic instructions, branches, loads/stores, etc. Monitoring instruction-level execution also considers the latency, overlapped with flash operations, thereby capturing very detailed \textit{per-request} and \textit{per-transaction} timing for both synchronous and asynchronous I/O services. In addition, we modify and integrate multicore power and area models \cite{mcpat} into the embedded CPU cores. This integration enables Amber to estimate the dynamic power and energy of firmware execution, based on the specific full-system environment being executed at runtime.

We also model an internal DRAM and its memory controller, which are connected to the CPU cores through the SSD system port. This internal DRAM and memory controller can capture detailed DRAM timing parameters, such as row precharge ($tRP$), row address to column address delay ($tRCD$), and CAS latency ($tCL$), considering all memory references made by the various firmware components.
Specifically, this memory model contains cached data, metadata and mapping information during the SSD simulation, which are dynamically updated by flash firmware. To measure the internal power consumption of an SSD, we also incorporate a DRAM power model \cite{drampower1,drampower2} in our DRAM module and controller. This integrated DRAM power model considers {\em all} different DDR memory states, including the power-down and self-refresh states, and also takes into account the  memory controller's open-page and close-page policies and all levels of bank-interleaving strategies.

As part of the storage complex, we modify and integrate a multi-channel and multi-way architecture \cite{simplessd} that can capture detailed timings, such as programming time ($tPROG$), flash memory island access time ($tR$), as well as latencies for data transfer ($tDMA$) and flash command operations. This integrated design allows Amber to accommodate highly-reconfigurable flash memory and controller models that can accurately mimic a diverse set of state-of-the-art flash technologies such as multi-level cell (MLC), triple-level cell (TLC), etc. Since this physical flash model has no built-in power model, we also modify and add a flash power measurement tool \cite{jung2016nandflashsim} into the storage complex's each flash package. This model can capture the dynamic power and energy consumption of data movements from the internal DRAMs to each package's row buffer (flash registers). In addition, it dynamically measures the actual flash access power consumed to load or store data between the buffer and flash die/plane.


\noindent \textbf{Firmware stack.} Figure \ref{fig:high_arch} shows Amber's firmware stack. At the top of this   firmware stack, HIL schedules I/O requests based on the queue protocol that the host storage interface defines. For example, HIL of h-type storage performs I/O scheduling based on first-in first-out (FIFO). However, it schedules I/O requests based on two I/O arbitration mechanisms for s-type storage, namely, round-robin (RR) and weighted round-robin (WRR). HIL fetches a host request from the device-level queue by communicating with a device controller that manages storage-side physical layer (PHY) and data movement. HIL then splits the request into multiple page-based internal requests by considering Amber's cache entry size, which is the same as the size of a super-page. The separated requests are cached into a DRAM model of the computation complex by the underlying \emph{internal cache layer} (ICL). ICL buffers/caches the data from the flash and/or host controller/driver. In cases where the data should be evicted due to a page replacement or flush command (coming from the host-side OS), ICL retrieves the corresponding data from the internal DRAM, composes per-page requests whose address indicates a super-page aligned LBA, and issues it to the underlying FTL. FTL then translates the request's LBA to \emph{super-page basis PPN} (S-PPN). If there is no available reserved block to allocate S-PPN, Amber's FTL performs GC (and wear-leveling) by considering the number of valid pages to migrate (e.g., Greedy \cite{bux2010performance}) and block's access time (e.g., Cost-benefit \cite{kawaguchi1995flash}). FTL then submits the super-page requests to the underlying \emph{flash interface layer} (FIL), which in turn schedules flash transactions and parallelizes I/O accesses across multiple channels and ways based on a parallelism method \cite{parallel} that the user defines. Further, we optimize ICL and FTL by being aware of flash internal parallelism to enhance overall performance, as will be explained later in Section \ref{subsec:optimization}.

Note that  all the software modules in Amber's firmware stack are highly {\em reconfigurable}, so that our simulation model can be incarnated as diverse storage devices under the full-system simulation environment. ICL can be configured as a fully-associative, set-associative, and direct-map cache; the number of ways and sets, cache entry size, and replacement policy (LRU, random, etc.) can all be reconfigured. Similarly, FTL can realize different types of mapping algorithms, such as block-level mapping, various hybrid mapping-algorithms, and pure page-level mapping. The request schedulers of FIL and flash controllers can also capture all possible parallelism combinations by taking into account the internal SSD resources such as a channel, way, die, and plane.

\noindent \textbf{Data transfer emulation.} In contrast to the existing SSD simulators that only capture latency or throughput based on timing calculator, Amber needs to handle the actual contents of all I/O requests between the host and storage. This data transfer emulation is necessary to be able to execute an OS and user-level applications on a  full-system environment. To this end, we model a DMA engine and integrate it into gem5, which transfers real data from the host's system memory to the internal DRAMs of Amber's SSD architecture model.
In practice, the host driver and/or controller for both h-type storage and s-type storage composes a pointer list whose each entry indicates a system memory page, as shown in Figure \ref{fig:high_arch}. The DMA engine that we implement in gem5 parses the pointer list, whose actual structure varies based on the interface protocol that a storage types defines. More detailed discussion on this will be presented later in Section \ref{sec:implementation}. The DMA engine then performs the data transfer from the host DRAM to the underlying storage.

\begin{figure*}[h!]
\captionsetup[subfigure]{captionskip=-3pt}
	\centering
	\begin{minipage}[b]{.49\linewidth}
		\subfloat[SATA.]{\label{fig:impl_sata}\rotatebox{0}{\includegraphics[width=0.49\linewidth]{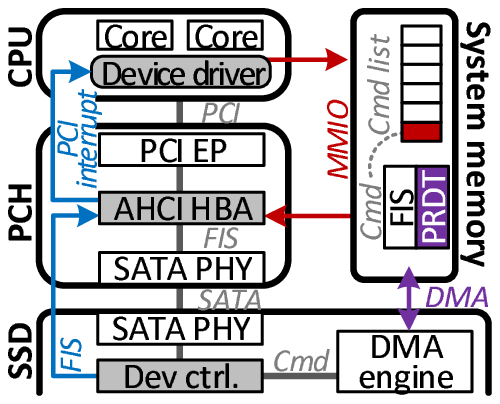}}}
		\hspace{1pt}
		\subfloat[UFS.]{\label{fig:impl_ufs}\rotatebox{0}{\includegraphics[width=0.49\linewidth]{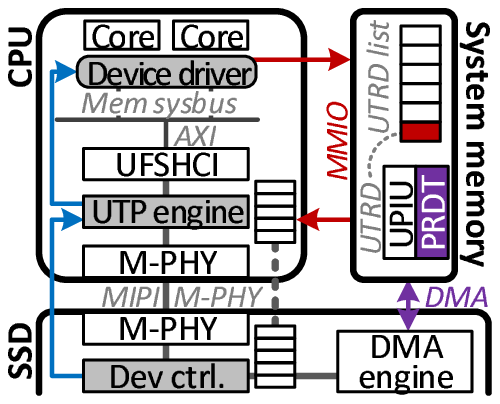}}}
		\caption{H-type storage\label{fig:impl_htype}.}
	\end{minipage}
	\begin{minipage}[b]{.49\linewidth}
		\subfloat[NVMe.]{\label{fig:impl_nvme}\rotatebox{0}{\includegraphics[width=0.50\linewidth]{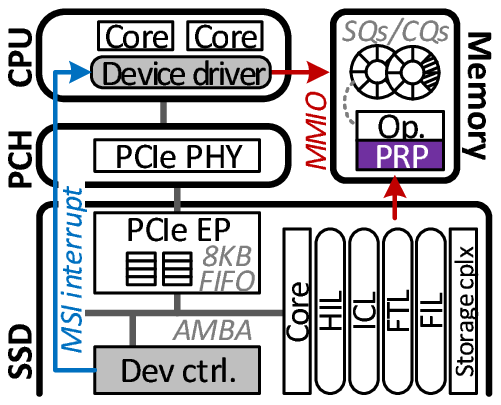}}}
		\hspace{1pt}
		\subfloat[OCSSD.]{\label{fig:impl_ocssd}\rotatebox{0}{\includegraphics[width=0.48\linewidth]{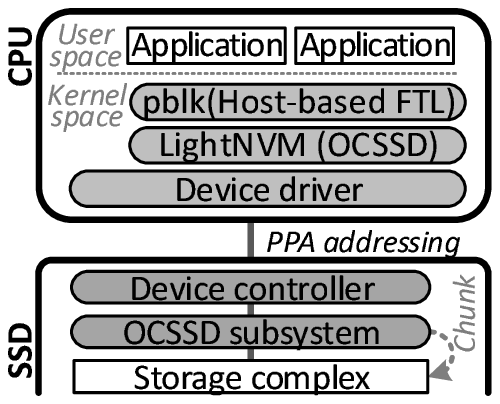}}}
		\caption{S-type storage\label{fig:impl_stype}.}
	\end{minipage}
	\vspace{-3pt}
\end{figure*}

One of the challenges behind this DMA engine implementation in a full system environment is that, the different CPU models in gem5 require different memory access timings and I/O service procedures for the software modules in the storage stack.
For example, a functional CPU model (i.e., \texttt{AtomicSimpleCPU}) requires to push/pull actual data at the very beginning of an I/O service (per request), but before the execution of any corresponding data communication activities.
This is because, such functional CPU employs a simplified DRAM model and does not have any specific timing for OS executions. However, all other timing CPU models, including in-order and out-of-order pipelined executions, use the details of the memory access timing for software emulations. Therefore, whenever the host and device controllers of s-type and h-type storage access the memory over the system bus, the DMA engine should be involved in handling the start and end of memory accesses. We define multiple DMA handlers that service queue (mapped to system memory) accesses, queue entry references, command compositions, and pointer list traversing activities, and register them with the gem5's timing simulation engine. For the content transmission, timing CPU models should be also involved in referring each system memory page access, transferring the page between the host and storage, and performing I/O completion based on MSI/MSI-X. While the functional CPU model just aggregates data transfer activities for each request into a single I/O task (as the pointer list contains many system memory pages to transfer) in timing CPU models, the DMA engine emulates finer-granular data transfers per I/O request by arbitrating the memory references of the device/host controllers and OS drivers as many times as the number of entries that the pointer list has.

\section{Details of Full-System Storage}
\label{sec:implementation}

\subsection{H-type Storage}

Figures \ref{fig:impl_sata} and \ref{fig:impl_ufs} illustrate our SSD implementations under full system environment for SATA and UFS, respectively. In SATA, the host's functional and timing CPU models are connected to processor controller hub (PCH), which integrates MCH and ICH together via front-side bus, whereas the CPU models directly communicate with the host controller in UFS.

\noindent \textbf{SATA modeling.} PCH attaches to the underlying SATA through a PCI endpoint by using PCI configuration spaces \cite{pci-config}. Under the PCI endpoint, a \emph{host block adapter} (HBA) generates a set of commands, pointer lists, queues and the corresponding data payloads. We implement HBA as SATA's host device controller in gem5, based on \textit{advanced host controller interface} (AHCI) specification \cite{ahci}. This HBA communicates with the OS device driver (e.g., ATA and the SCSI libraries \cite{ata-lib, scsi-driver}), which exposes the underlying SSD simulation model to all upper-layer OS components as an actual storage volume.
Specifically, HBA exposes two major sets of registers, which are mapped to the system memory: one for a \textit{command list} and another for \textit{communication information} \cite{ahci}. The command list contains 32 entries, which is related to SATA's native command queue (NCQ) management, while the communication information handles data transfer-related packets, referred to as \emph{frame information structure} (FIS). Each command in the command list includes a pointer list that contains system memory page addresses; in SATA, the pointer list is callled \emph{physical region descriptor table} (PRDT) \cite{prdt}. Through HBA's memory-mapped register sets, OS drivers can manage the order of I/O requests and set all the corresponding FIS and payload data. HBA then fetches a command and FISs from the register sets, and issues the I/O request through SATA PHY. In our implementation, SATA PHY exists for both full system simulator and storage simulator. Thus, the device controller parses the issued SATA requests, and pass them to HIL. During this phase, the DMA engine in gem5 retrieves all the target page addresses in the system memory through PRDT and emulates the data transfers, while I/O completions and interrupts (for a host CPU) are managed by both the device controller and HBA.

\noindent \textbf{UFS modeling.}  Even though UFS employs h-type storage protocol management, which is very similar to the aforementioned SATA, the UFS datapath between the host CPU and the host controller is slightly different from that of SATA.
Since UFS is designed for handheld computing, the host controller resides in CPU as an SoC, which is directly connected to the memory system bus. As shown in Figure \ref{fig:impl_ufs}, we implement the host controller based on \emph{UFS transport protocol} (UTP) \cite{ufs}, referred to as \emph{UTP engine}. UTP engine is connected to the system bus, called \textit{advanced extensible interface} (AXI), instead of PCIe \cite{pcie}. The host CPU can communicate with the UTP engine through \emph{UFS host controller interface} (UFSHCI) \cite{ufshci}. UTP engine and UFSHCI are functionally equivalent to SATA HBAs and PCI endpoint, respectively. The physical layer of UFS is defined by M-PHY \cite{mphy}, which exists both on the SoC side and on the storage side. To address the issue of different frequency domains between the UTP engine and the underlying device controller, we also implement a small-sized FIFO queue within those two controllers. The queuing and data transfer methods are very similar to SATA's NCQ and PRDT. Also, similar to SATA, the UTP engine exposes a set of registers via memory-mapped address spaces to the host CPU. Thus, OS drivers (e.g., OS and UFS host controller drivers \cite{ufshcd}) can issue a command, referred to as \emph{UTP transfer request descriptor} (UTRD) \cite{ufshci}, over the command list which has 32 queue FIFO queue entries, similar to SATA/NCQ. Each queue entry contains PRDT and two \emph{UFS protocol information unit} (UPIU) \cite{ufs} for request and response. Once the I/O request is issued by UTP engine, the device controller, parses all commands, data payloads, and communication information, and then exposes them to HIL. The DMA engine emulates data transfers, and the corresponding I/O completion and interrupt are managed by both the device controller and UTP engine.




\subsection{S-type Storage}

Figure \ref{fig:impl_stype} shows the s-type storage details that Amber implements in a full-system environment. NVMe and OCSSD employ the PCIe physical interface as opposed to the ICH interface, and simplify the host-side datapath by removing the host controller. PCH's multiple PCIe lanes are connected to a PCIe endpoint (that exists on the SSD-side). We model the endpoint based on Xilinx's FPGA Gen3 Integrated Block for PCIe. Our endpoint employs inbound and outbound 8KB FIFO queues, which convert PCIe PHY to SSD on-chip interconnect system bus, called \textit{advanced microcontroller bus architecture} (AMBA) \cite{amba}, by modeling a AXI4-stream interconnect core that can connect multiple master/slave AMBA ports.

\noindent \textbf{NVMe modeling.}
NVMe controller parses the PCIe and NVMe packets as the device-side controller, and exposes the corresponding information to HIL, so that HIL can handle the I/O requests by collaborating with other firmware modules such as ICL, FTL and FIL. We leverage a protocol/queue management module \cite{simplessd-x}, but significantly modify gem5's system barbus, DMA engine. We also implement the device controller to support all mandatory NVMe commands and some optional features (such as NVMe namespace management and a scatter gather list (SGL)) for supporting diverse demands of the host-side software. The communication strategies between NVMe controller and OS driver (e.g., NVMe driver) are functionally similar to h-type storage, but the queue mechanism of NVMe (and OCSSD) is quite different from that of h-type storage. As NVMe removes the host controller, instead of h-type storage's per-controller queue, OS can create 65536 rich queues, each containing up to 65536 entries. Each queue logic is coupled with a \emph{submission queue} (SQ) and a \emph{completion queue} (CQ), which are synchronized between the OS drivers and the NVMe controller. Specifically, OS drivers issue an I/O service, compose 64 bytes NVMe request and put the request into an SQ entry. This 64 bytes request includes an op-code (i.e., NVMe command), metadata (e.g., namespace), and a pointer list. In NVMe, the pointer list is implemented by \emph{physical region page} (PRP) List \cite{nvme}, which can contain more than 512 system memory pages (4KB). In our implementation, we also offer a conventional scatter-gather list (SGL) \cite{sgl} in cases where a different version of OS drivers needs SGL. When the service is completed, an event of I/O completion is delivered by the NVMe controller to the OS drivers through 16 bytes CQ request. The queue states between the NVMe driver and controller are synchronized over MSI and a set of registers (called \emph{doorbell}) that we implemented. During this time, the DMA engine emulates the transfer of all relevant data (presented by PRP/SGL) and NVMe packets.

\noindent \textbf{OCSSD modeling.}
Since OCSSD overrides the NVMe interface/protocol to expose physical flash addresses to the host, the datapath and hardware architecture (e.g., PCIe endpoint) are the same as in NVMe. The difference between NVMe and OCSSD is that OCSSD unboxes SSD's storage complex and exposes the low-level flash page addresses to the host software, whereas NVMe still keeps all flash managements on the SSD side as a kind of \emph{active} device. Specifically, OCSSD allows OS and user software to tune the underlying SSD subsystems by employing FTL and ICL on the host side, and it considers the underlying SSD as a \emph{passive} device. To enable OCSSD in a full-system simulation environment, we first implement both OCSSD 1.2 \cite{ocssd1} and 2.0 interfaces \cite{ocssd2}, and enable \texttt{pblk} \cite{pblk} and \texttt{lightNVM} \cite{lightnvm} modules in gem5's Linux kernel, as a host-side FTL and OCSSD driver, respectively. \texttt{LightNVM} is tightly coupled with the existing OS's NVMe driver to handle NVMe-like I/O services, but also handles the overridden OCSSD command and feature management.
We also implement an OCSSD subsystem on the SSD-side, which leverages NVMe's device controller and HIL, but supports all the features required for OCSSD specification, such as chunk (similar to physical block) and other information that the host-side OS requires (e.g., memory latency data and erase count information). Since \texttt{pblk} manages flash addresses underneath file systems, the OCSSD subsystem in SSD simulations disables FTL and ICL from the datapath of the firmware stack.


\subsection{Flash Firmware Optimizations}
\label{subsec:optimization}

\noindent \textbf{Internal caching.} ICL can cover most of cache associativities and page replacement schemes, but the read performance of the simulated SSD can still be behind that of a real device. While writes can be immediately serviced from the internal DRAMs, reads should be serviced from the taget flash, and therefore, performance can be directly impacted by different physical data layouts and various levels of internal parallelism \cite{chen2011essential,jung2012evaluation}. To address this, ICL performs a \textit{parallelism-aware readahead} that loads multiple super-pages, existing across all physical dies, but having no resource conflict at flash-level, in advance. This readahead is activated only if multiple incoming requests exhibit a high degree of data locality. ICL keeps track of multiple start offsets and page lengths of the requests, which are cached in DRAM, and increases a frequency counter if the incoming requests are sequentially accessed right after the addresses of the previous ones (but no cache hit). When the frequency counter becomes greater than a threshold, which is reconfigurable, ICL performs the readahead.

\begin{table}
\centering
\begin{adjustbox}{width=\linewidth}
\begin{tabular}{|c|c|c|}
\hline
 & \textbf{PC platform} & \textbf{Mobile platform} \\ \hline
\textbf{CPU name} & Intel i7-4790K & NVIDIA Jetson TX2 \\ \hline
\textbf{ISA} & X86 & ARM v8 \\ \hline
\textbf{Core number} & 4 & 4 \\ \hline
\textbf{Frequency} & 4.4GHz & 2GHz \\ \hline
\textbf{L1D cache} & \begin{tabular}[c]{@{}c@{}}private, 32KB, 8-way \end{tabular} & \begin{tabular}[c]{@{}c@{}}private, 32KB \end{tabular} \\ \hline
\textbf{L1I cache} & \begin{tabular}[c]{@{}c@{}}private, 32KB, 8-way \end{tabular} & \begin{tabular}[c]{@{}c@{}}private, 48KB \end{tabular} \\ \hline
\textbf{L2 cache} & \begin{tabular}[c]{@{}c@{}}private, 256KB, 8-way \end{tabular} & \begin{tabular}[c]{@{}c@{}}shared, 2MB \end{tabular} \\ \hline
\textbf{L3 cache} & \begin{tabular}[c]{@{}c@{}}shared, 8MB, 16-way \end{tabular} & N/A \\ \hline
\textbf{Memory} & DDR4-2400, 2 channel & LPDDR4-3733, 1 channel \\ \hline
\end{tabular}
\end{adjustbox}
\vspace{2pt}
\caption{Gem5 system configurations. \label{tab:sys_config}}
\vspace{2pt}
\end{table}

\begin{table}[]
	\vspace{5pt}
	\centering
	\begin{adjustbox}{width=\linewidth}
		\begin{tabular}{|c|c|c|c|c|c|}
			\hline
			& \begin{tabular}[c]{@{}c@{}}Avg. read\\ length (KB)\end{tabular} &
			\begin{tabular}[c]{@{}c@{}}Avg. write\\ length (KB)\end{tabular} & \begin{tabular}[c]{@{}c@{}}Read\\ ratio (\%)\end{tabular} & \begin{tabular}[c]{@{}c@{}}Random\\ read (\%)\end{tabular} & \begin{tabular}[c]{@{}c@{}}Random\\ write (\%)\end{tabular} \\ \hline
			\textbf{Authentication Server (\textit{24HR})} & 10.3 & 8.1 & 10 & 97 & 47 \\ \hline
			\textbf{Back End SQL Server (\textit{24HRS})} & 106.2 & 11.7 & 18 & 92 & 43 \\ \hline
			\textbf{MSN Storage metadata (\textit{CFS})} & 8.7 & 12.6 & 74 & 94 & 94 \\ \hline
			\textbf{MSN Storage FS (\textit{MSNFS})} & 10.7 & 11.2 & 67 & 98 & 98 \\ \hline
			\textbf{Display Ads Payload (\textit{DAP})} & 62.1 & 97.2 & 56 & 3 & 84 \\ \hline
		\end{tabular}
	\end{adjustbox}
	\vspace{2pt}
	\caption{Workload characteristics.\label{tab:workload}}
	\vspace{-5pt}
\end{table}

\begin{figure*}
	\centering
	\subfloat[Sequential read.]{\label{fig:bw_sr}\rotatebox{0}{\includegraphics[width=0.24\linewidth]{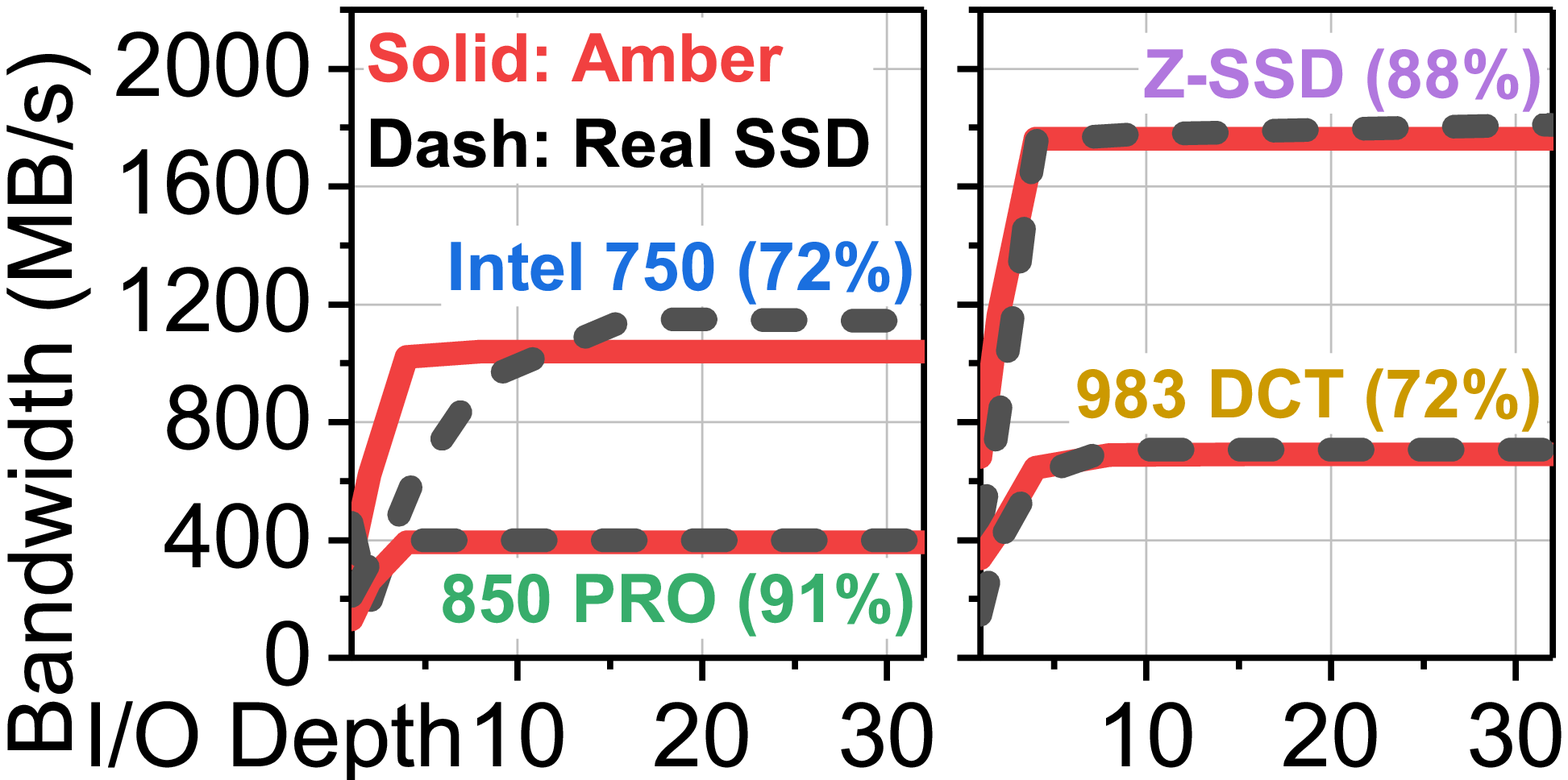}\hspace{5pt}}}
	\subfloat[Random read.]{\label{fig:bw_rr}\rotatebox{0}{\includegraphics[width=0.24\linewidth]{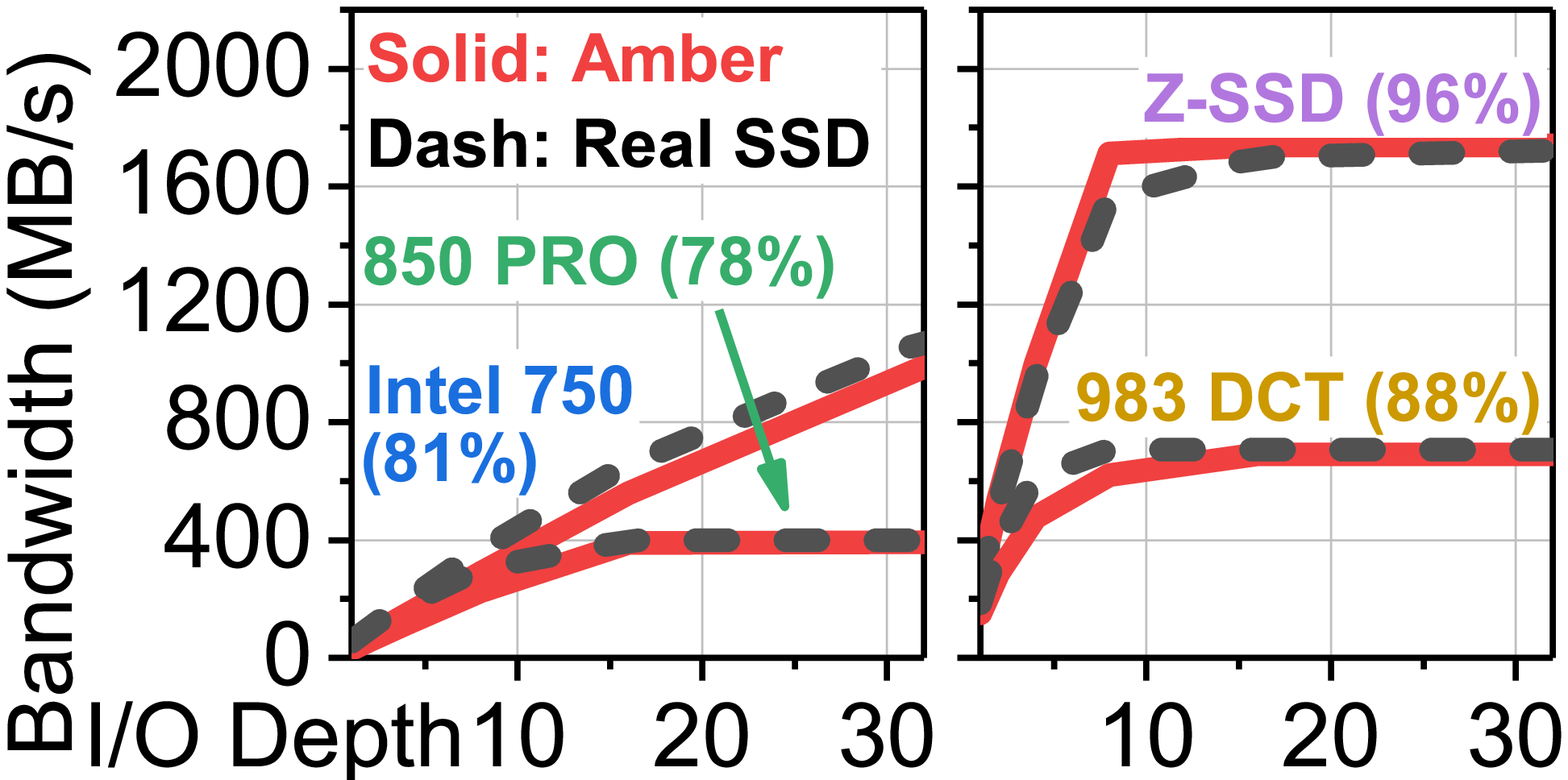}\hspace{5pt}}}
	\subfloat[Sequential write.]{\label{fig:bw_sw}\rotatebox{0}{\includegraphics[width=0.24\linewidth]{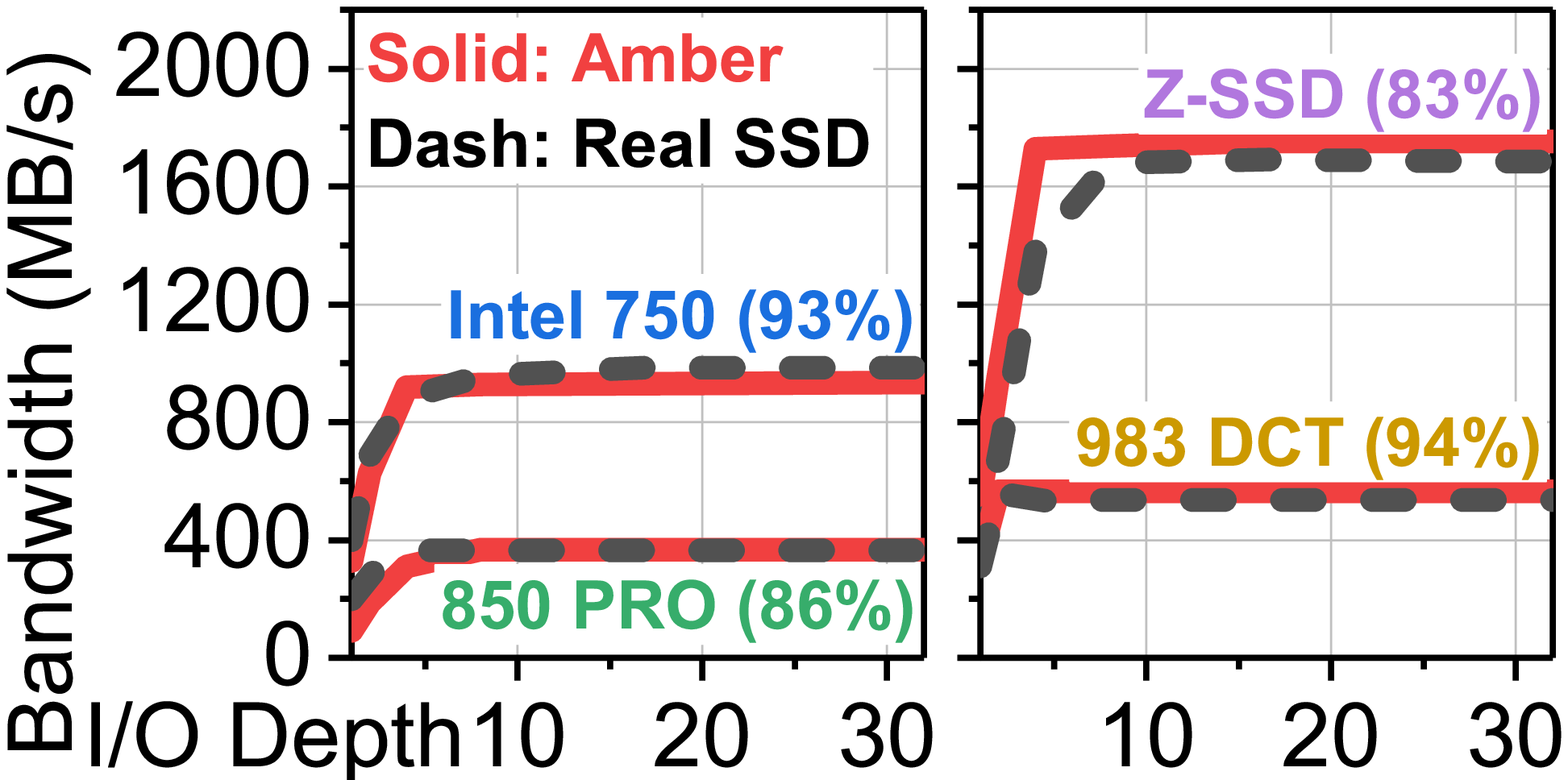}\hspace{5pt}}}
	\subfloat[Random write.]{\label{fig:bw_rw}\rotatebox{0}{\includegraphics[width=0.24\linewidth]{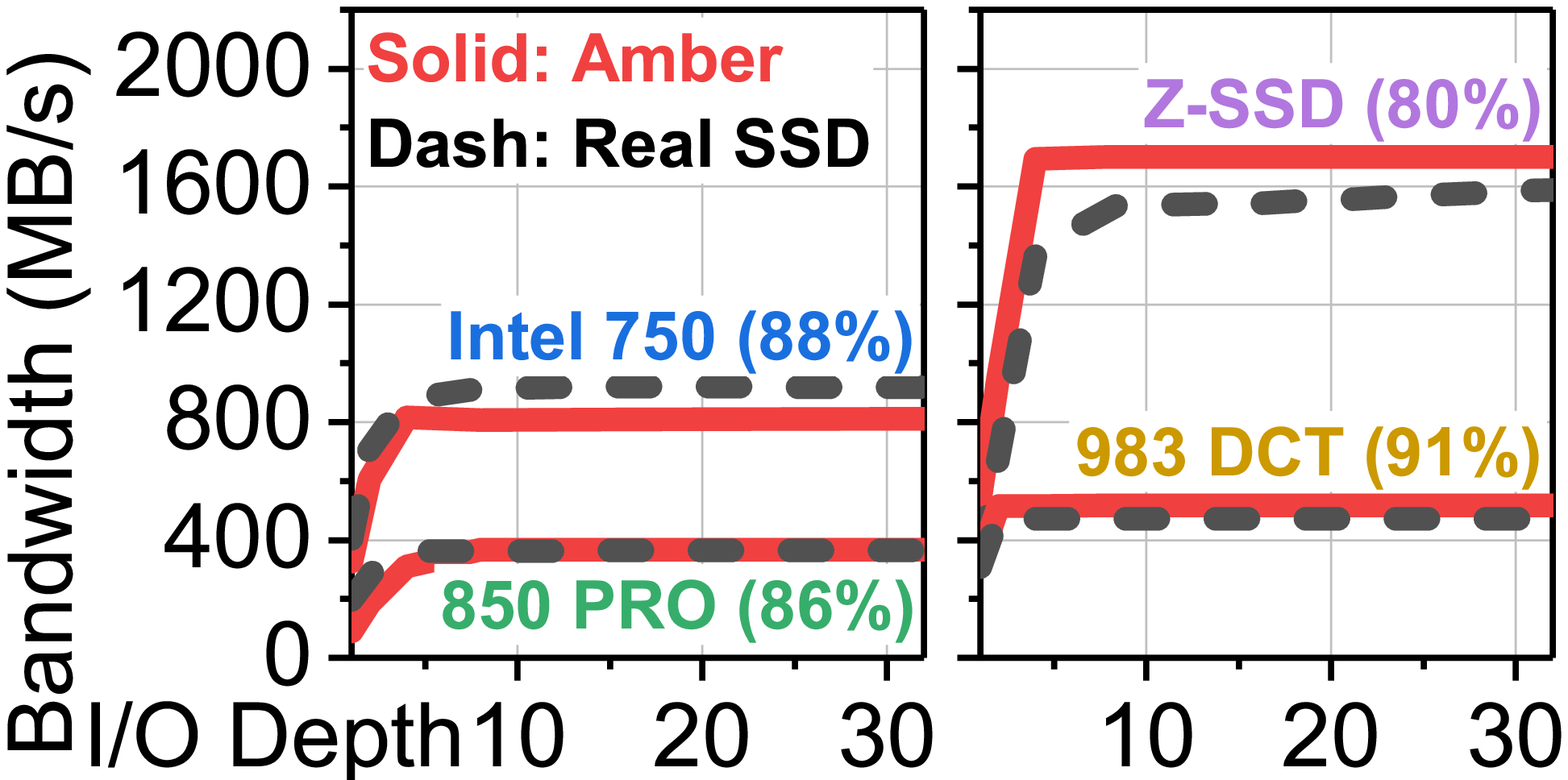}\hspace{5pt}}}
	\caption{Bandwidth trend and accuracy comparisons for real devices' and Amber's simulation results.}
	\vspace{-4pt}
	\label{fig:val_bw}
\end{figure*}

\begin{figure*}
	\centering
	\subfloat[Sequential read.]{\label{fig:lat_sr}\rotatebox{0}{\includegraphics[width=0.24\linewidth]{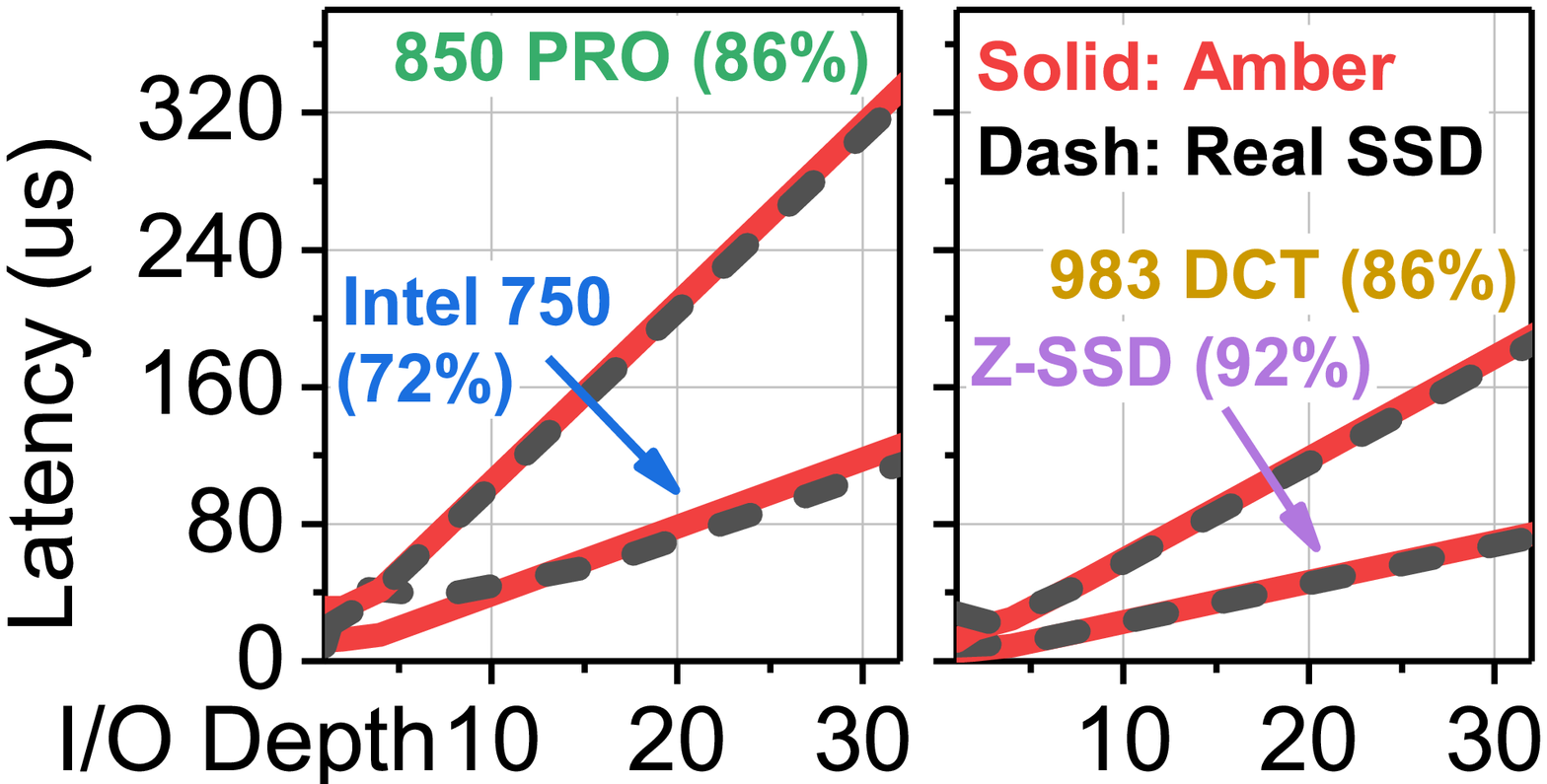}\hspace{5pt}}}
	\subfloat[Random read.]{\label{fig:lat_rr}\rotatebox{0}{\includegraphics[width=0.24\linewidth]{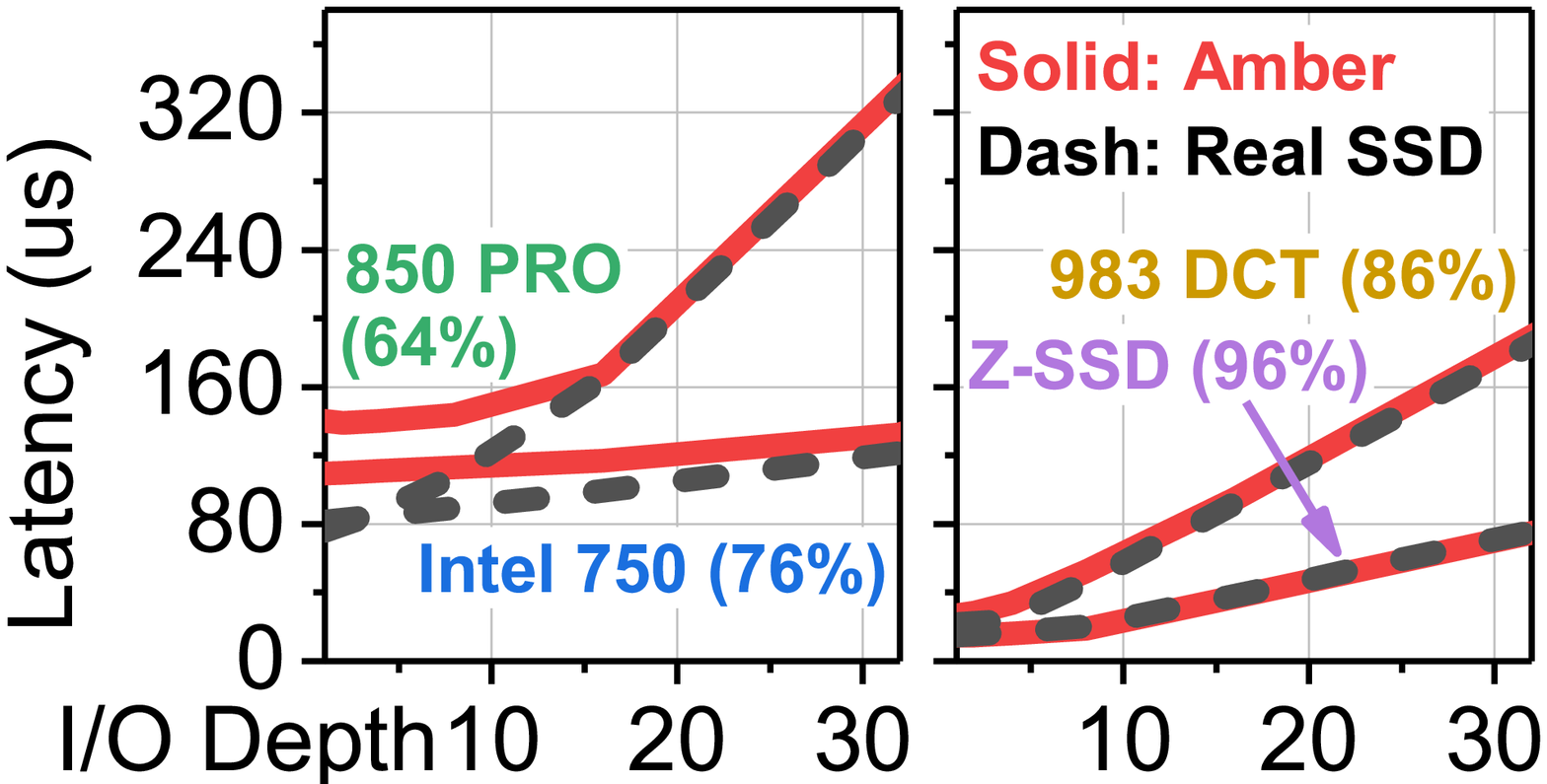}\hspace{5pt}}}
	\subfloat[Sequential write.]{\label{fig:lat_sw}\rotatebox{0}{\includegraphics[width=0.24\linewidth]{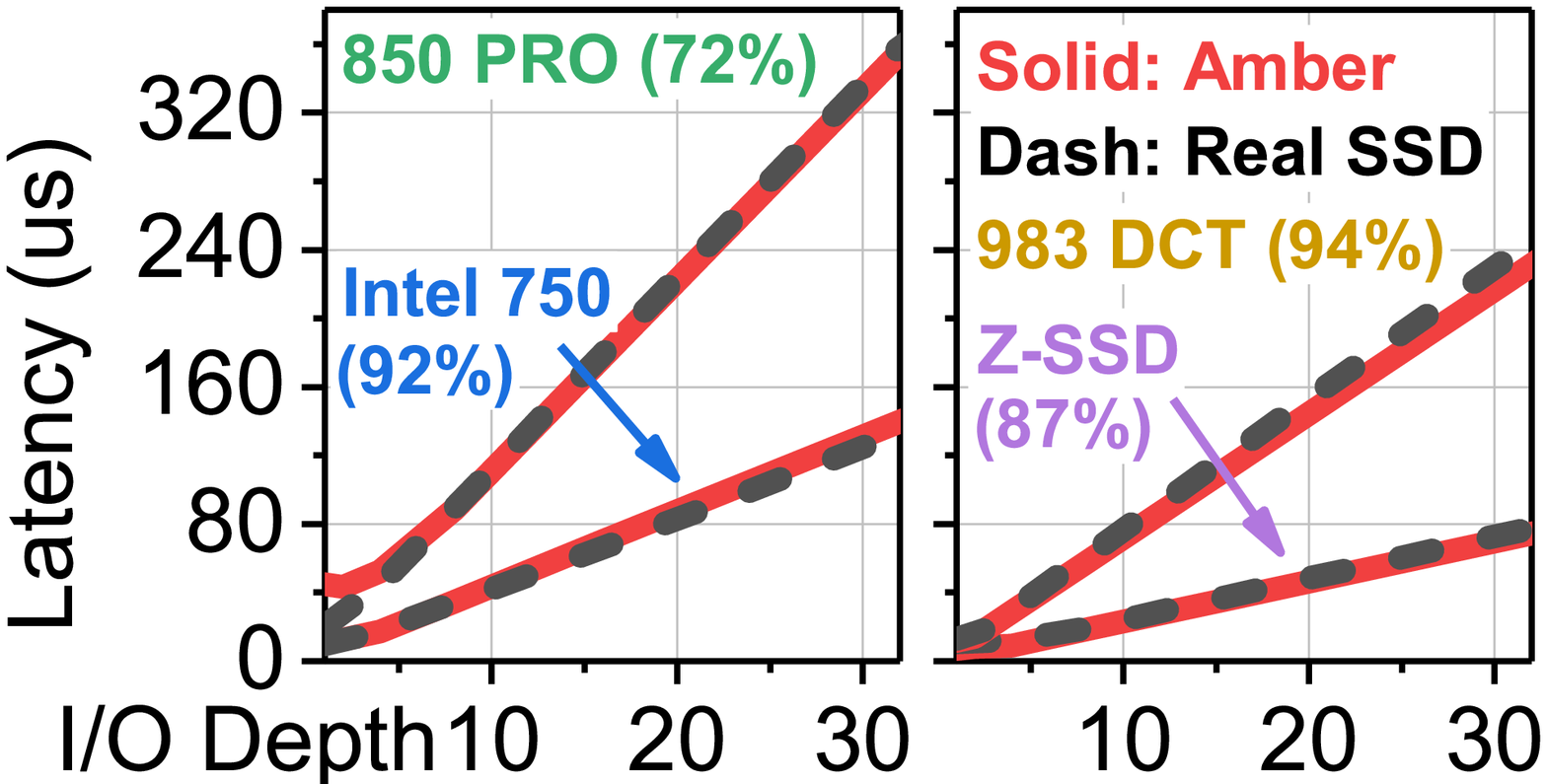}\hspace{5pt}}}
	\subfloat[Random write.]{\label{fig:lat_rw}\rotatebox{0}{\includegraphics[width=0.24\linewidth]{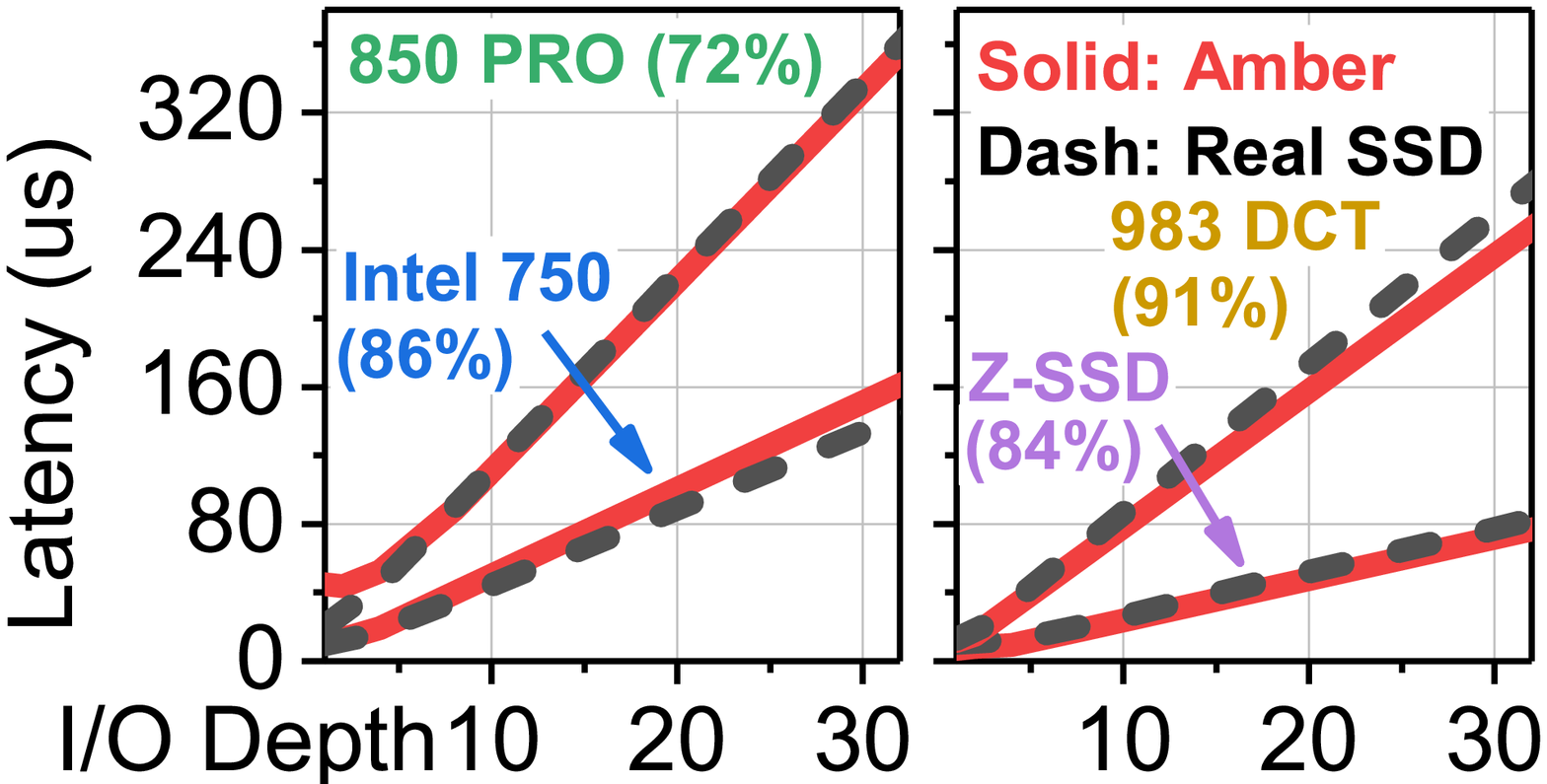}\hspace{5pt}}}
	\caption{Latency trend and accuracy comparisons for real devices' and Amber's simulation results.}
	\vspace{-2pt}
	\label{fig:val_lat}
\end{figure*}

\noindent \textbf{FTL mapping.} Even though a super-page based mapping algorithm increases the level of internal parallelism, thereby achieving higher bandwidth, we observed that small random writes can significantly degrade overall performance. This is because small-sized writes introduce many read-and-modify operations, which give a burden on the storage complex. Specifically, if the target cache line of ICL to evict/flush is smaller than a super-page unit, ICL needs to first access other physical pages associated with the target super-page, and then write the pages (but still as a super-page) to storage complex, which in turn introduces read-and-write intermixed I/O workloads thereby making more resource conflicts.
A challenge behind this optimization is that, when ICL writes a small-sized page from its eviction, the actual address of the target page should be updated by FTL, due to the erasure-before-write characteristic. If FTL maps the addresses based on the physical page unit, it needs to search all page table entries to find out other pages, which can be spread across all other channels (or ways). Thus, if there is a small-sized write, FTL selectively remaps the target page (but exists in a same channel or way) within the corresponding super-page by employing a super-page based hashmap.

\section{Evaluation}
\label{sec:evaluation}

\subsection{Methodologies}
\noindent \textbf{Simulation configurations.} We configure the SSDs with 60 flash packages whose programming latency vary from 413 us to 1.8 ms, and the range of read latency is between 57 us and 94 us. This latency variation fundamentally exists in most MLC flash, due to the material-level nature of incremental step purse programming (ISPP) \cite{ispp}. The SSD interconnection between storage complex and computation complex is implemented via AXI (250MHz) and AMBA (250MHz), and the interface linking the underlying flash is ONFi 3 (333MHz). 3 ARMv8 CPU cores are used for executing our SSD firmware. As  default, we use a page-level mapping for FTL, and ICL is configured as a fully-associative cache. We model DDR3L for internal DRAM, and for SSD interfaces, we use SATA 3.0, NVMe 1.2.1, UFS 2.1 and OCSSD 2.0. The default configuration of storage complex (e.g., channels, packages, dies, etc.) is the same as that of the real device we used in Table \ref{tab:hwconfig} of Section \ref{sec:highlevelview}. To explore a full space of simulator evaluations, we also apply different storage complex configurations similar to the other testbeds that we used in Section \ref{sec:validation}. Table \ref{tab:sys_config} tabulates the host system configuration and timing parameters for gem5.

\noindent \textbf{Workloads.} The important characteristic and corresponding descriptions of our workloads \cite{tracetracker} are listed by Table \ref{tab:workload}.
While most of the workloads exhibit small-sized requests (8$\sim$10 KB), the average request sizes of \emph{24HRS}(W2) and \emph{MSNFS}(W5) are 28KB and 74KB, respectively.
All the evaluations presented in this section are performed by \emph{running actual storage applications at user-level} with a full system stack, instead of replaying or generating traces within the storage simulator.


\subsection{Validation}
\label{sec:validation}
\noindent \textbf{Real devices.} In addition to the real device (Intel 750), we prepare three more real SSD devices and compare their performance with simulation results: i) Samsung 850 PRO (H-type), ii) Samsung Z-SSD prototype (S-type), and iii) Samsung 983 DCT SSD prototype (S-type). We evaluated other NVMe and SATA devices, but their performance significantly fluctuate/untenable and is far away from the above, and therefore, we carefully select those four devices as comparison target testbeds. 850 PRO is composed by multiple MLC flash over 8 interconnectors. 983DCT is similar to 850 PRO, but it supports the multi-stream feature (newly added in NVMe). Z-SSD uses NVMe interface and protocol, but their backend media is replaced with new flash that supports 3us and 100us I/O latency for read and write, respectively \cite{isscc-paper}. Note that \emph{all configurations of our Amber that we evaluate for this validation will be available to download with the simulation source codes.}

\noindent \textbf{Bandwidth comparison.} Figure \ref{fig:val_bw} compares the user-level performance trend of four different SSDs simulated by Amber (solid lines) and real devices (dashed lines). The figure also includes the average accuracy of our simulation for each real target device. One can observe from these plots that, for all micro-benchmark evaluations (e.g., reads, writes, sequential and random patterns) the Amber's bandwidth curve exhibits a very similar trend compared to that of the real devices, as the I/O queue depth increases. Specifically, the average accuracy of Amber's simulated bandwidth for Intel 750 ranges from 88\% and 93\% for sequential and random writes, and for reads, it also exhibits 72\% $\sim$ 81\% accuracy while closely following the bandwidth curve of the real device. Even though real devices exhibit different curve patterns with different bandwidth levels, the accuracy of Amber is in the range of 72\% $\sim$ 96\%.

\noindent \textbf{Latency comparison.} Figure \ref{fig:val_lat} shows the latency trend and accuracy of simulated results compared to each real target device when varying I/O queue depth. As shown in the figure, the user-level latency curve simulated by Amber almost overlaps with that of the real device under all micro-benchmarks as the I/O depth increases, and the accuracy of simulation ranges between 64\% and 94\% for all real devices. The difference between simulation results and real device latency (especially Intel 750 and 850 PRO) is relatively more notable at a low I/O queue depth (1 $\sim$ 8), compared to the cases with higher queue depths. We believe that this is because Intel 750 and 850 PRO have several optimizations, which operate at a specific I/O pattern. Even though the optimization techniques such as caching data with battery-backed DRAM \cite{stockdale2004high} or dumping them to the underlying flash by only using fastest pages of MLC are not published in a public domain, the varying and unsustainable performance of such devices with a specific I/O pattern are reported in various publicly available articles \cite{kim2016nvmedirect,nikkel2016nvm}.

\begin{figure*}
	\centering
	\subfloat[Sequential read.]{\label{fig:bs_bw_sr}\rotatebox{0}{\includegraphics[width=0.24\linewidth]{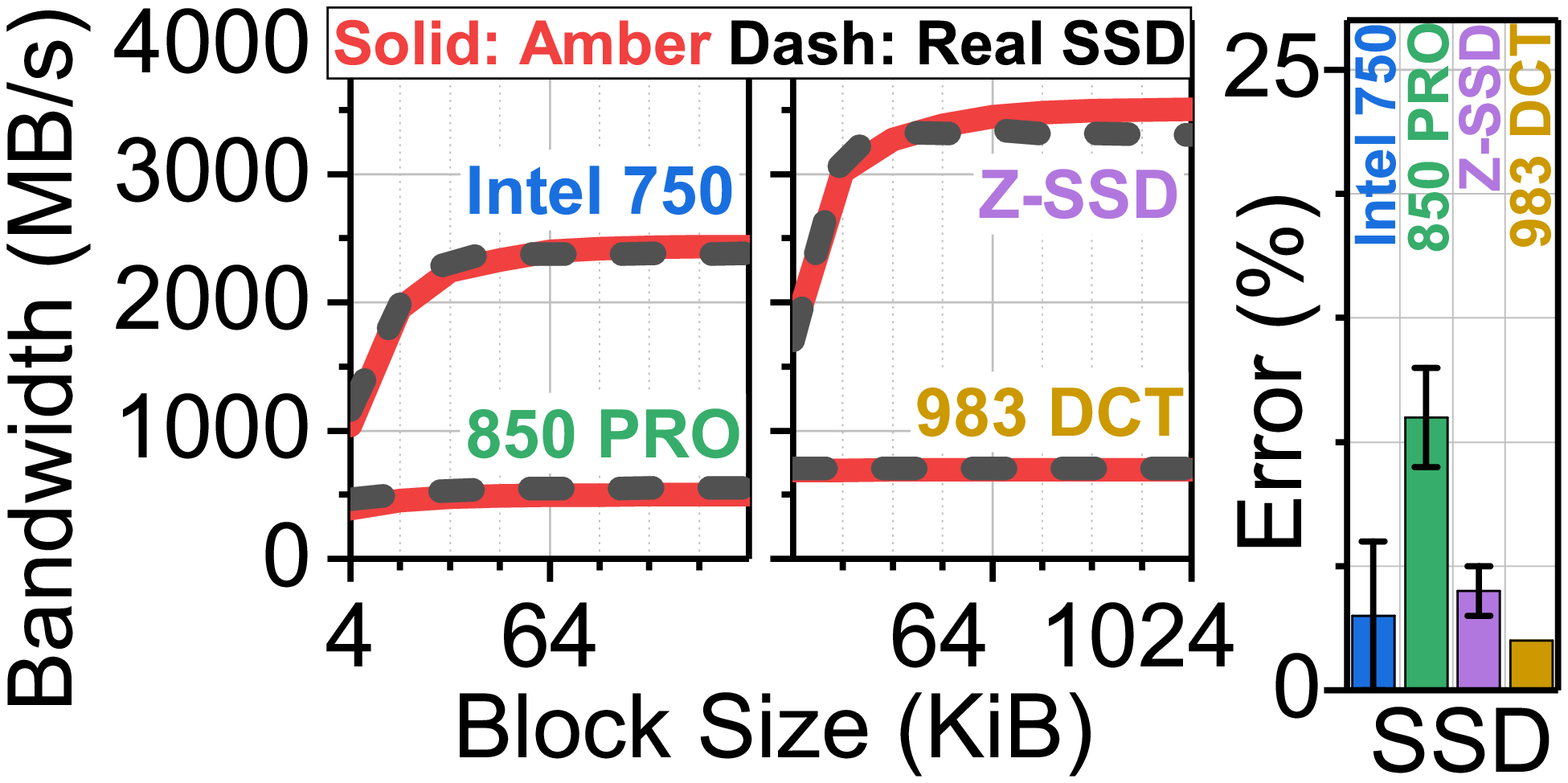}\hspace{5pt}}}
	\subfloat[Random read.]{\label{fig:bs_bw_rr}\rotatebox{0}{\includegraphics[width=0.24\linewidth]{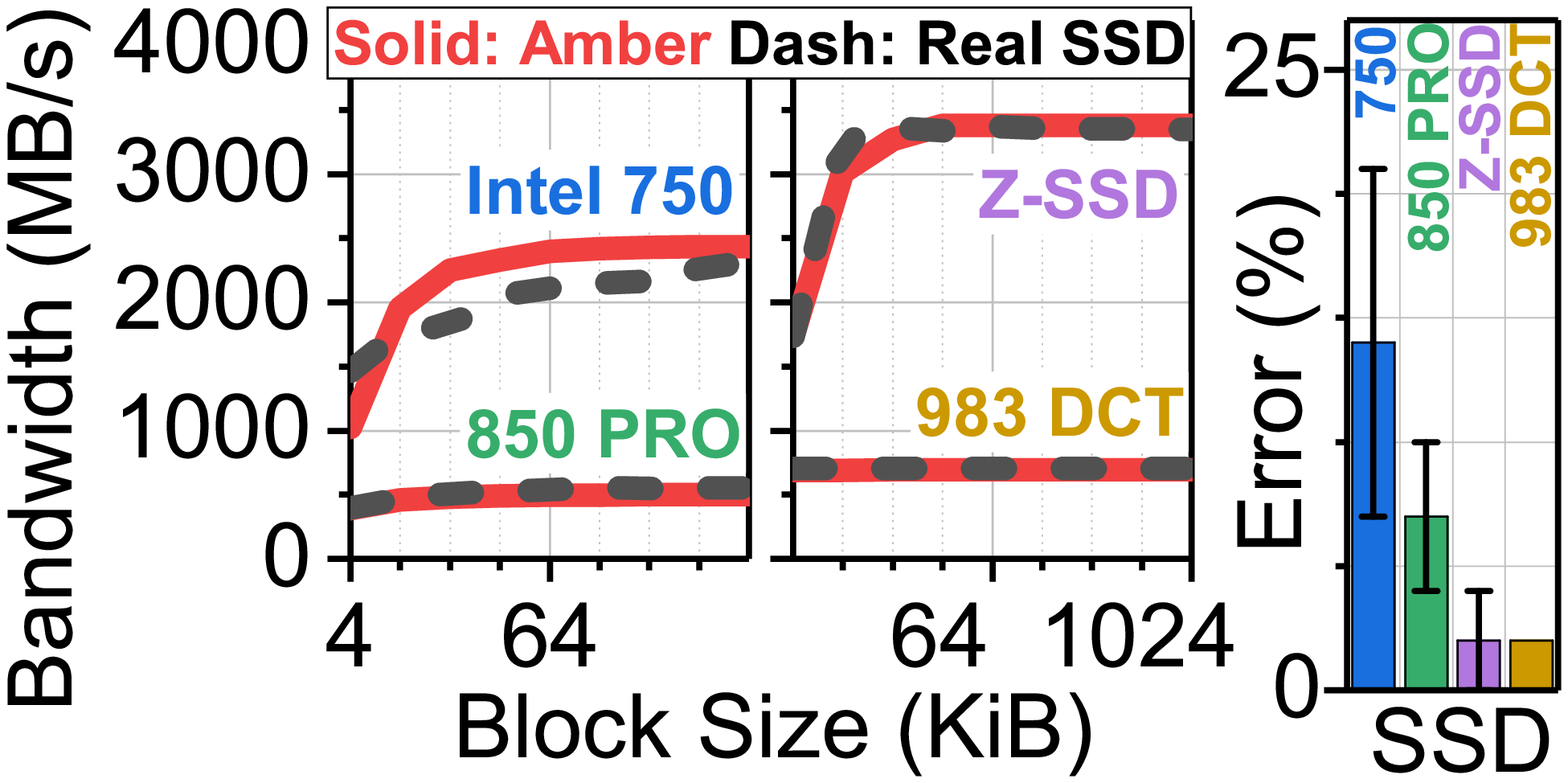}\hspace{5pt}}}
	\subfloat[Sequential write.]{\label{fig:bs_bw_sw}\rotatebox{0}{\includegraphics[width=0.24\linewidth]{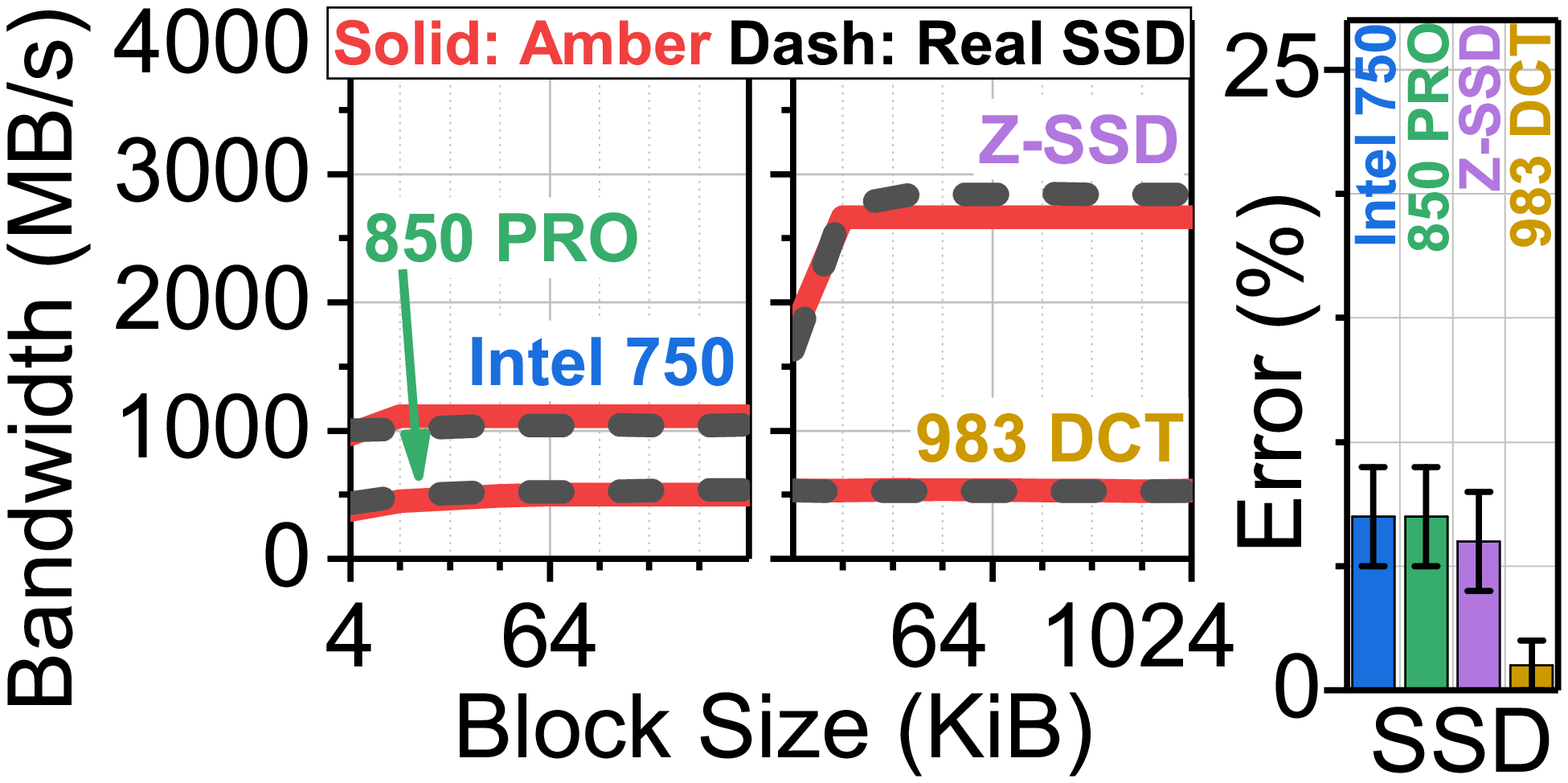}\hspace{5pt}}}
	\subfloat[Random write.]{\label{fig:bs_bw_rw}\rotatebox{0}{\includegraphics[width=0.24\linewidth]{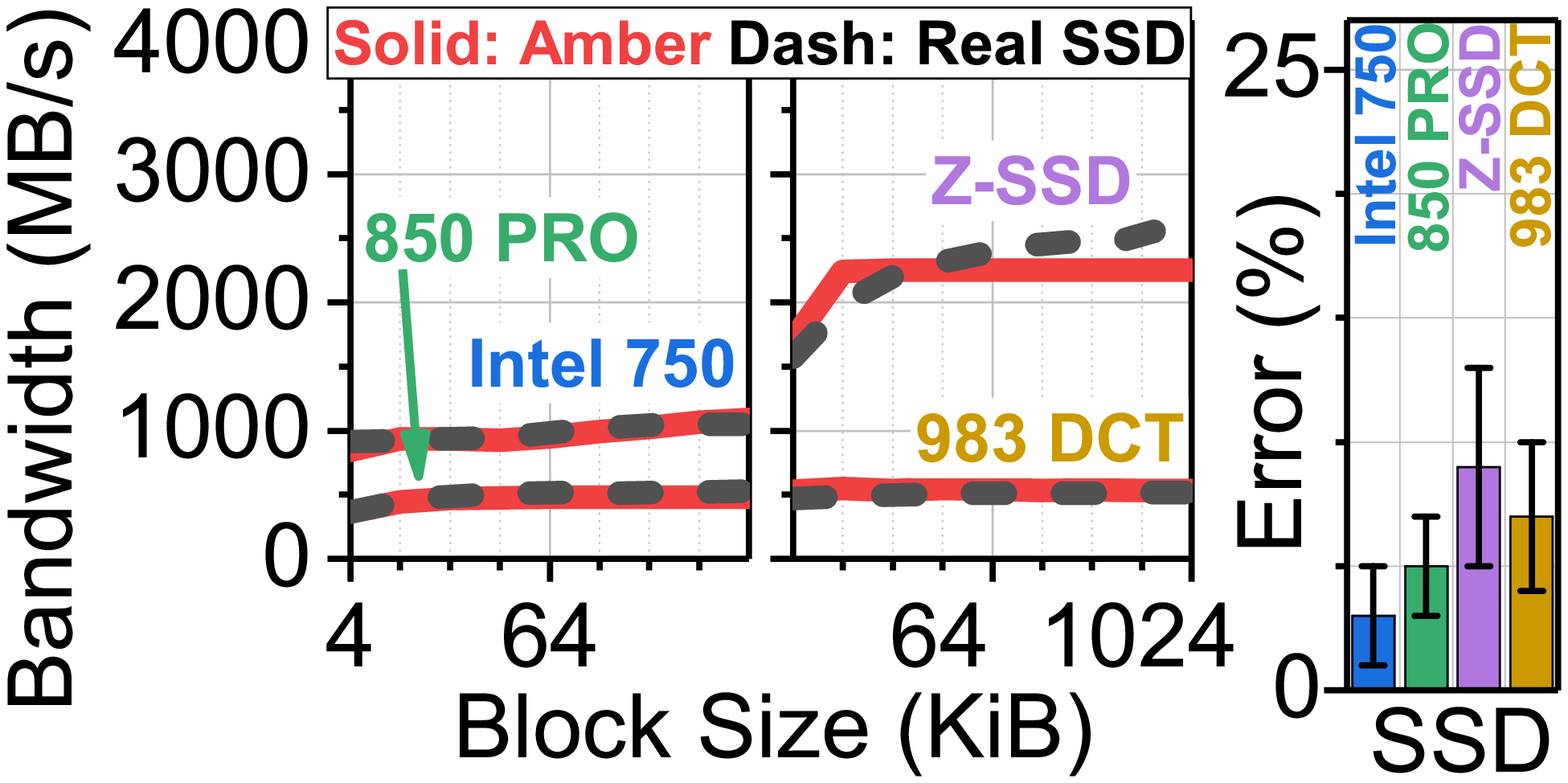}\hspace{5pt}}}
	\caption{Performance validation with different block sizes ranging from 4KB to 1024KB.}
	\label{fig:bs_bw}
	\vspace{3pt}
\end{figure*}

\noindent \textbf{Validation with different block sizes.} Figure \ref{fig:bs_bw} shows bandwidths of the real device and Amber. In this evaluation, we increase the block size from 4 KB to 1024 KB. The figure also includes error rate results for each device and the corresponding range (from minimum errors to maximum error rates), as an evaluation summary. The error range is calculated by $|(Perf_{real} - Perf_{sim})| / Perf_{real}$, where $Perf_{real}$ and $Perf_{sim}$ are the real devices' performance and Amber's simulation results, respectively. As shown in the figure, the performance curve from simulation is very similar to the performance curves of all real devices with the tests of varying block sizes. While Amber simulation results catch up the trend of real device performance, the error rates (measured at user-level of the host) are in still reasonable range of 6\% for all sizes of I/O request blocks across all SSDs. In these sensitivity tests in terms of different block sizes, Intel 750 shows higher error rates than those of other real devices (14\% in random read, on average). We believe that this is because of the internal optimizations that we mentioned in the latency comparisons.

\noindent \textbf{Over-provisioning.} Note that all of above validations are performed after fully filling the target storage spaces with sequential writes for each test set (e.g., STEADY-STATE). Even though we perform all evaluations with the steady-state, the write performance can vary based on the ratio of over-provision (OP). Thus, we evaluate the write performance with different block sizes (4KB $\sim$ 1024KB) by reducing the OP ratio from 20\% (default as same with Intel 750's OP rate) to 5\%, and the results are plotted in Figure \ref{fig:optest}. To create a worst-case scenario as a stress test, we randomly write data for the entire space into the steady-state SSDs, which means that the amount of written data is 2$\times$ greater than the actual target volume. The simulation results with 15\% and 10\% and 5\% OP rates show significant performance drops (by 87.9\%, 62.1\%, and 33.7\%, respectively). This is mainly because of too many invocations, which lead to frequent migrations of data over different flash packages, thereby introducing not only long I/O latencies but also frequent resource conflicts.

\subsection{Operating System Impacts}

\begin{figure*}
	\centering
	\begin{minipage}[b]{.51\linewidth}

	\begin{minipage}[b]{.28\linewidth}
	\includegraphics[width=\linewidth]{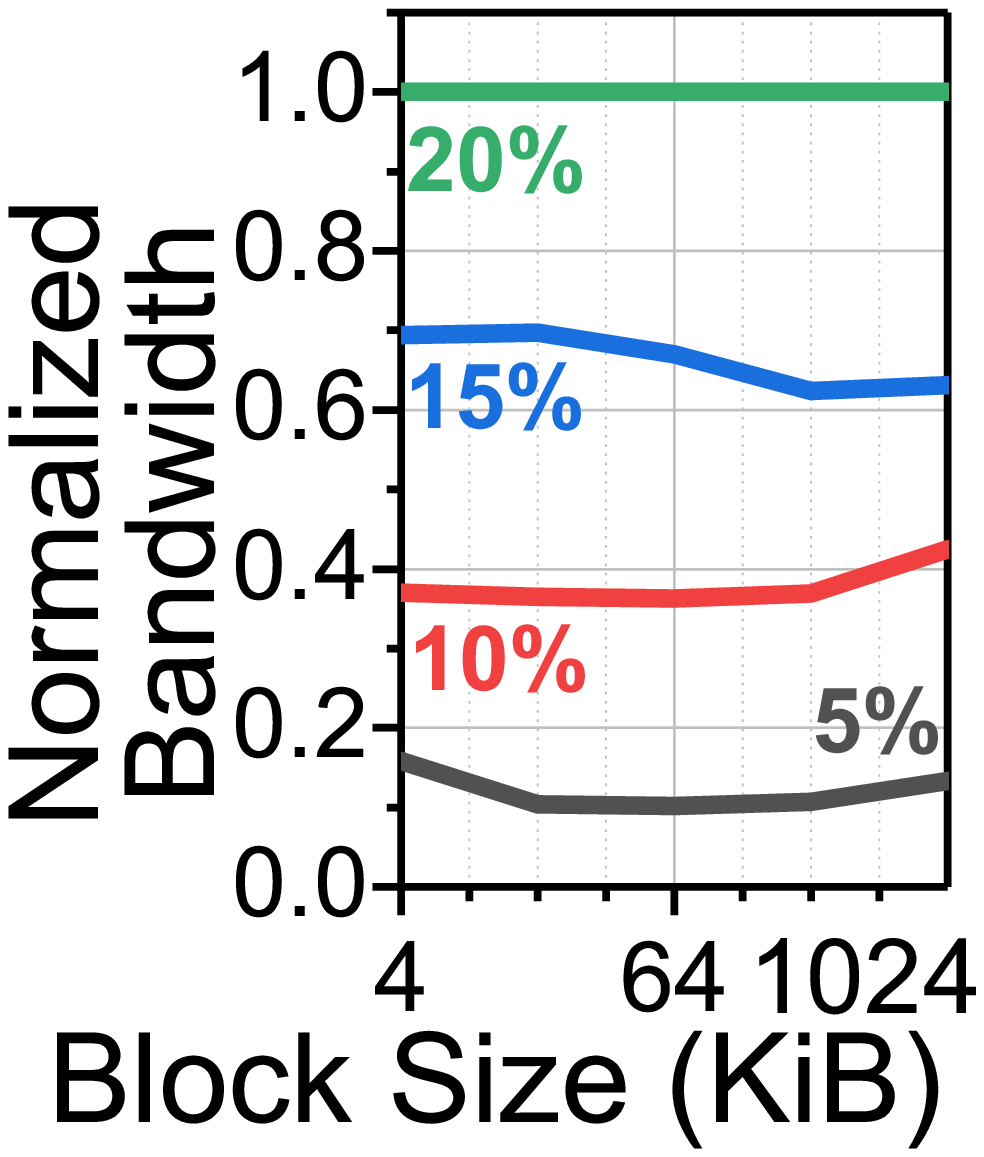}
	\caption{OP.}
	\label{fig:optest}
	\end{minipage}
	\begin{minipage}[b]{.68\linewidth}
	\includegraphics[width=\linewidth]{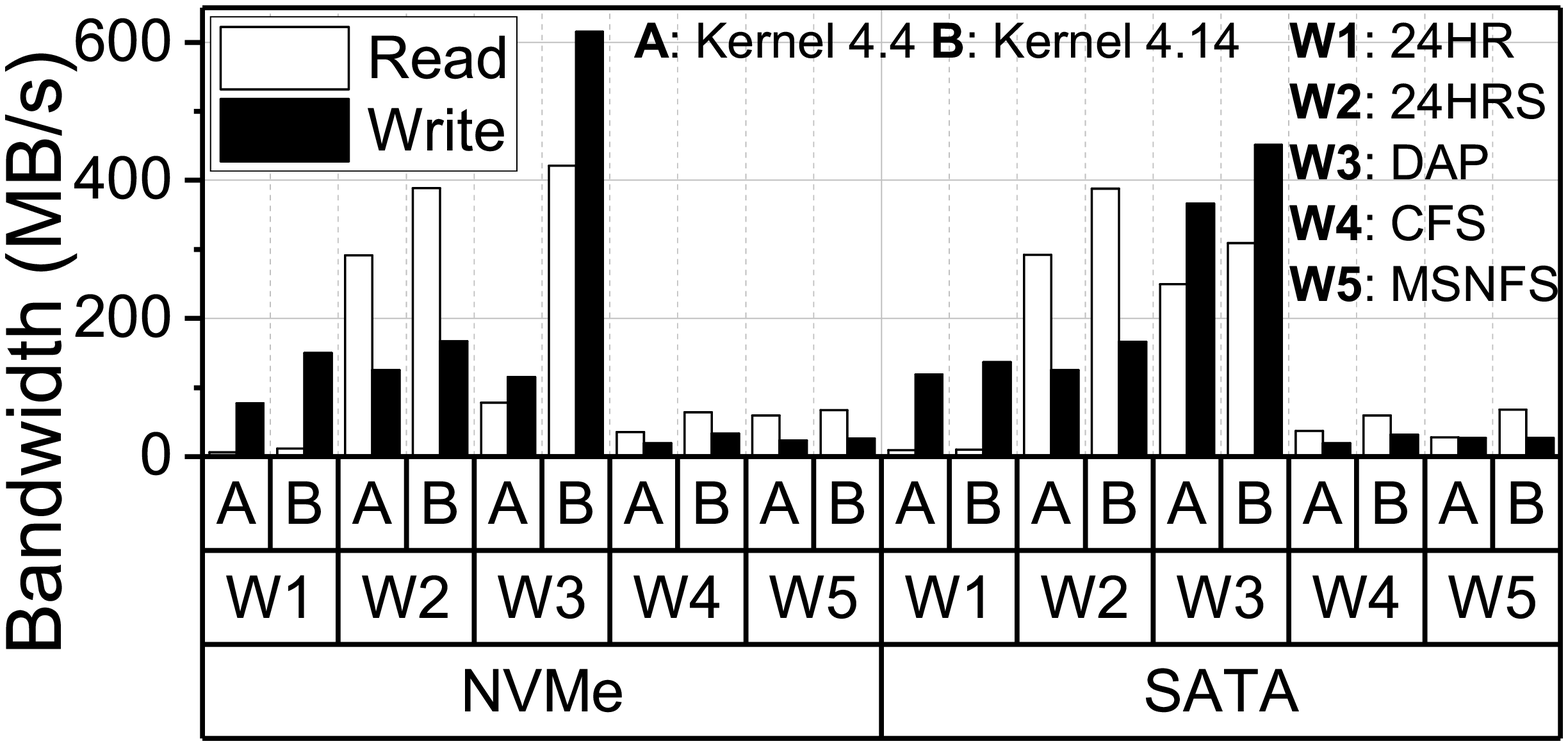}
	\caption{Performance impacts on OS.}
	\label{fig:workload}
	\end{minipage}

	\end{minipage}
	\begin{minipage}[b]{.47\linewidth}

	\centering
	\vspace{-10pt}
	\subfloat[Performance.]{\label{fig:eval_mobile_bw}\rotatebox{0}{\includegraphics[width=0.42\linewidth]{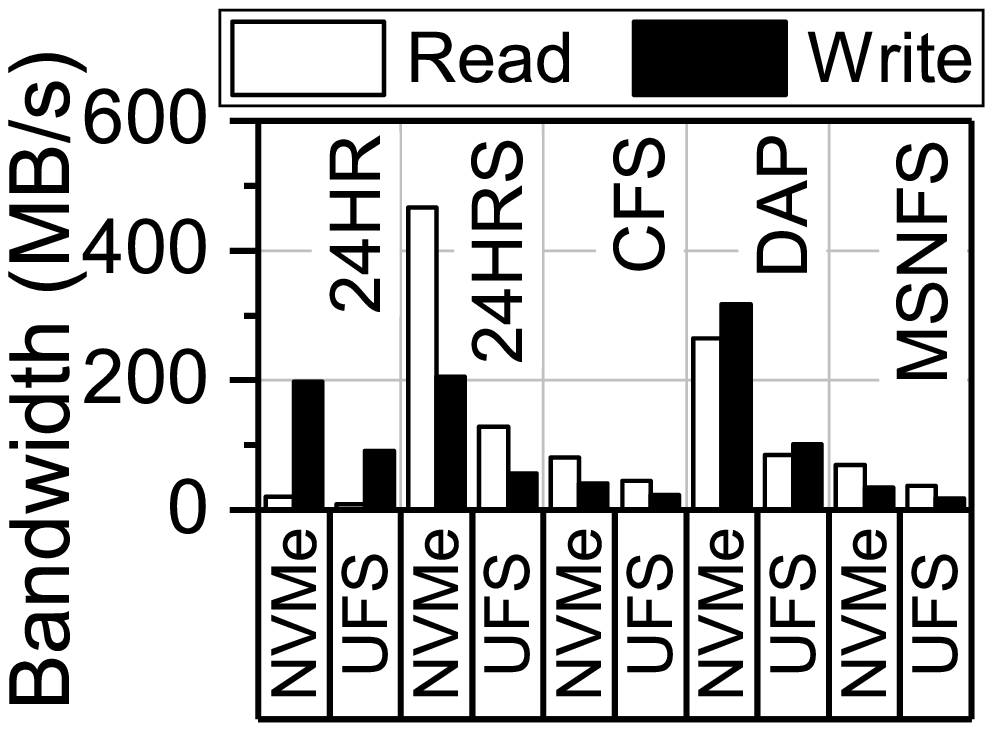}\hspace{5pt}}}
	\subfloat[Power.]{\label{fig:eval_mobile_energy}\rotatebox{0}{\includegraphics[width=0.27\linewidth]{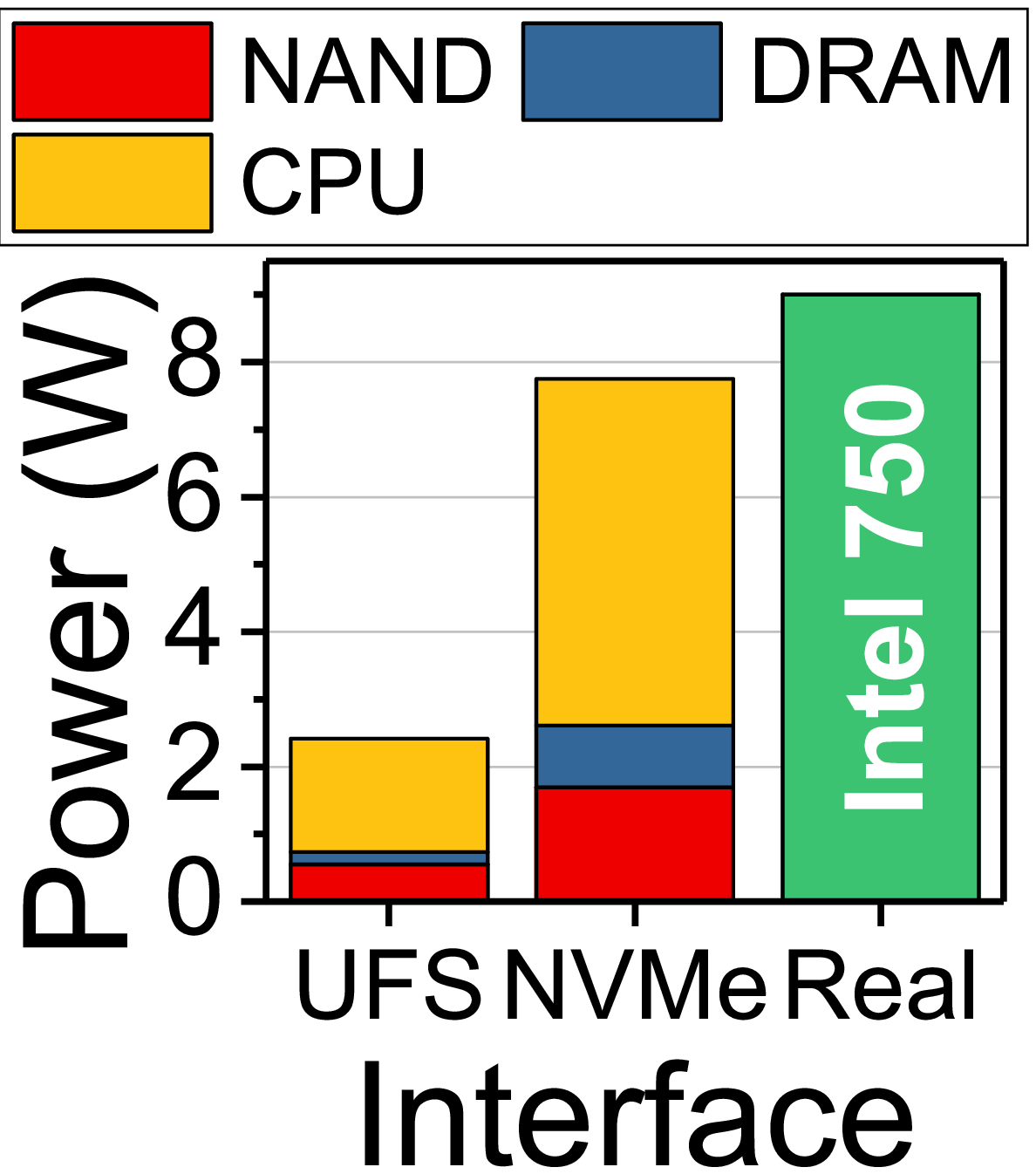}\hspace{5pt}}}
	\subfloat[Instruction.]{\label{fig:eval_mobile_inst}\rotatebox{0}{\includegraphics[width=0.27\linewidth]{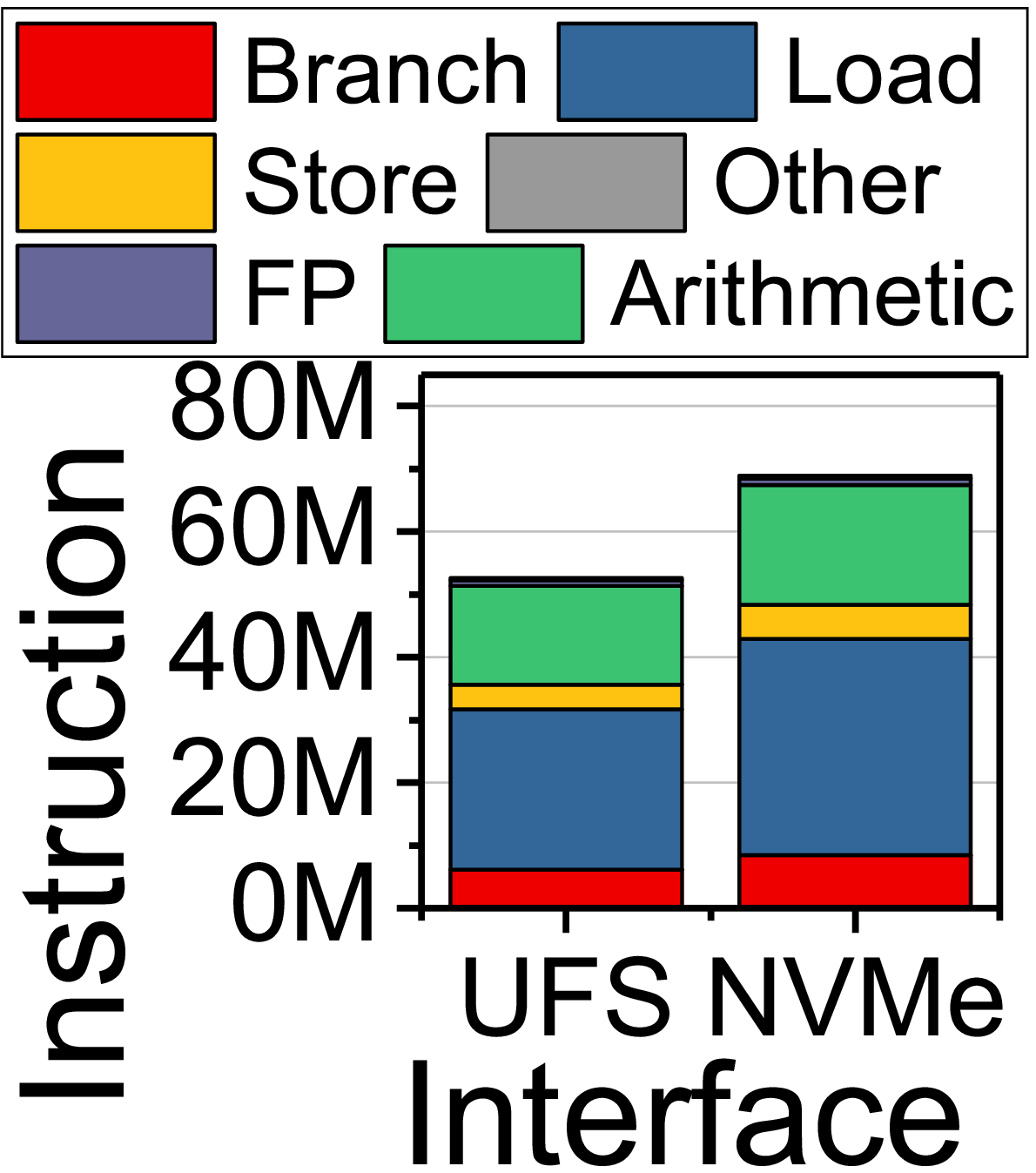}\hspace{5pt}}}
	\caption{General computing vs. handheld computing}
	\label{fig:mobile}

	\end{minipage}
	\vspace{-2pt}
\end{figure*}

Figure \ref{fig:workload} shows the performance of real workloads (executed at a user-level) when employing different versions of OS (4.4 and 4.14). Interestingly, the read and write performances of Kernel 4.4 are worse than that of Kernel 4.14 by 63\% and 69\%, on average, respectively. The major difference between those two versions of Linux kernel from the viewpoint of storage stack is the disk scheduler. The newer version of Kernel (4.14) employs a refined Budget Fair Queueing (BFQ) \cite{bfq}, which assigns a budget to each time slice for scheduling I/O requests by considering the number of sectors (i.e., request length). This refined BFQ is different from the original BFQ as it employs per-process queues optimized for SSDs. BFS also uses a unified mechanism to merge incoming requests to improve the performance through sequential access patterns. In contrast, the older version of Kernel (4.4) changes the disk scheduler to Completely Fair Queuing (CFQ) \cite{cfq}, which removes an anticipatory scheduling mechanism \cite{iyer2001}, and allows the other kernel scheduler to dispatch the I/O requests thereby improving throughput. Even though one would expect the newer version of Kernel that reflects the latest features of NVMe and SATA, compared BFQ, CFQ is not able to generate enough I/O requests to saturate the SSD performance and consumes CPU cycles in I/O scheduling.
To dig deeper into these OS-bound issues that prevent the system from serving enough I/O requests, we increase Amber's host-side CPU clock frequency from 2 GHz to 8 GHz by employing the fastest SSD (Z-SSD), and the results are given in Figure \ref{fig:cpu_test}. Thanks to the new flash technology, the performance at the device-level now reaches at 4.3 GB/s. However, Kernel execution (User-level) and protocol management (Interface-level) with 2 GHz CPU degrades the performance by 41\% compared to the device-level performance. When we run the same Kernel with a higher frequency (8GHz), the user-level performance is enhanced by 12\%. Note that Kernel (with 8GHz operations) still slashes device-level performance by 29\%, but we believe that future efforts that tune the Kernel and scheduler to make them better aware of the storage stack can reduce the performance loss more, and Amber can help such efforts as a research vehicle with the full functionality of holistic system simulation.

\subsection{Handheld vs. General Computing}
Figure \ref{fig:eval_mobile_bw} compares the user-level performance of a mobile system and that of a personal computing device, which use UFS and NVMe protocols, respectively. One can observe from this figure that NVMe shows 1.81 times better performance than UFS. Even though the performance of NVMe exceeds that of UFS, for \emph{24HR}, \emph{CFS} and \emph{MSNFS} workloads. The difference between them is less than 15\%. This is because of the low computing power in the mobile system, as described in the previous section; the low power mobile cores cannot generate enough I/O requests to enjoy all potential benefits of NVMe performance. As one would expect that several handheld devices such as tablets will have higher computing power in the future, NVMe might be a better solution to take off the storage accesses from the critical path. However, to this end, NVMe devices also require a significant technology enhancement. Specifically, Figures \ref{fig:eval_mobile_energy} and \ref{fig:eval_mobile_inst} breakdown the power and instruction sets for the SSD executions, respectively. The CPU that exists at the SSD-side is the most power hungry component. Considering the power budget of handheld computing, it requires significantly optimization of the underlying SSD's power usage.
As shown in Figure \ref{fig:eval_mobile_inst}, load and store are most dominant instructions, which account for 60\% of the total executions executed. We also observe that NVMe can execute more instructions than UFS within a same time period (by 5.45 times). This is because the core that handles NVMe queue should be involved everytime the doorbell rings. Note that, as shown in the figure, the power that Amber reports for NVMe, including internal CPU, flash-backend (NAND) and DRAM, is similar to that of the NVMe real device (Intel 750). While the power simulated by UFS is around 2W, the majority of such power is consumed by the internal CPU, and therefore, it can be reduced to hundreds $m$W with hardware automation in the mobile storage.

\subsection{Passive vs. Active Storage}

\begin{figure*}
	\centering
	\begin{minipage}[b]{.24\linewidth}
		\stackunder[8pt]{\includegraphics[width=1\linewidth]{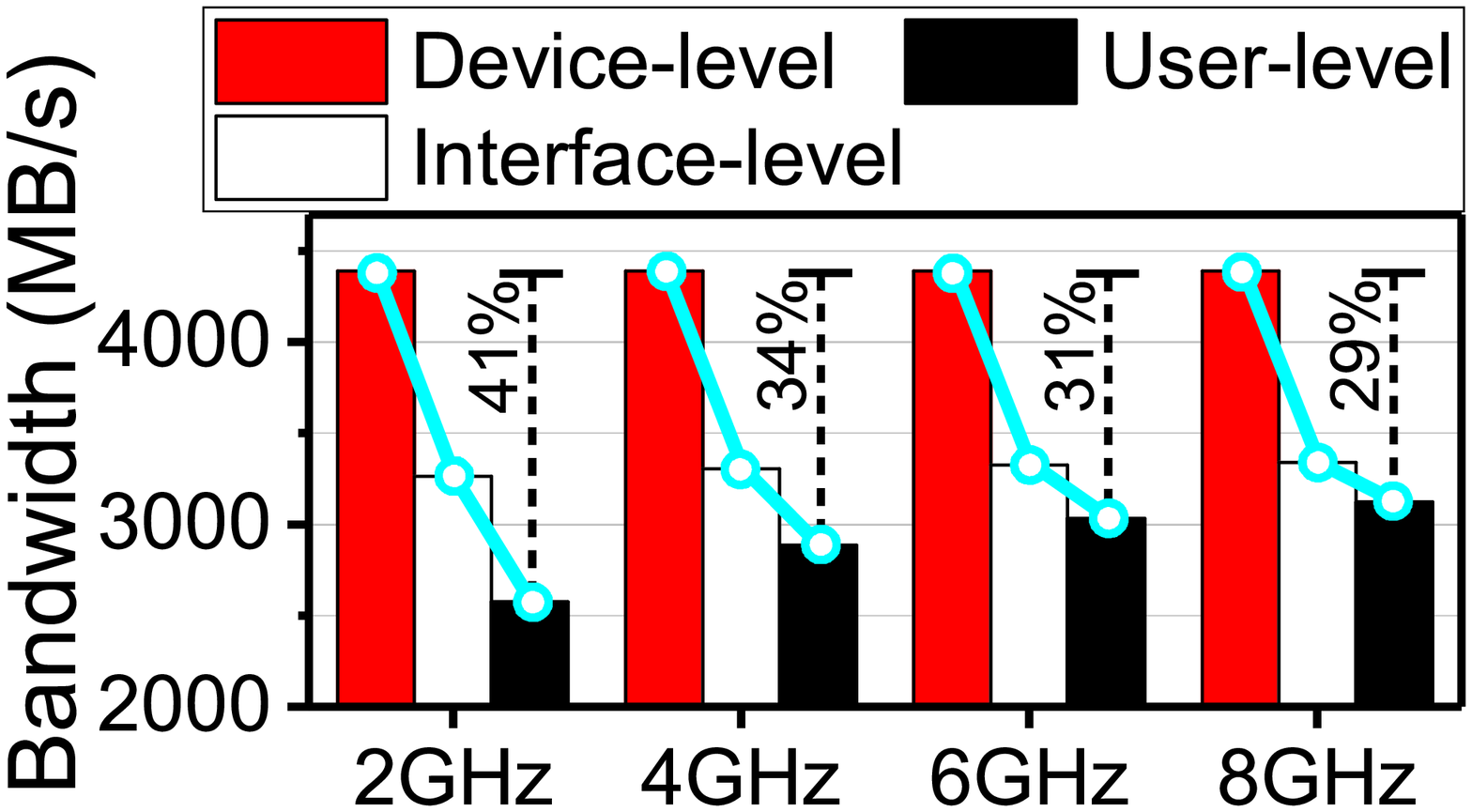}}{ }
		\vspace{-10pt}
		\caption{\label{fig:cpu_test}Performance with CPU frequency changes.}
	\end{minipage}
	\begin{minipage}[b]{.74\linewidth}
	\subfloat[Overall performance.]{\label{fig:ocssd_bw}\rotatebox{0}{\includegraphics[width=0.33\linewidth]{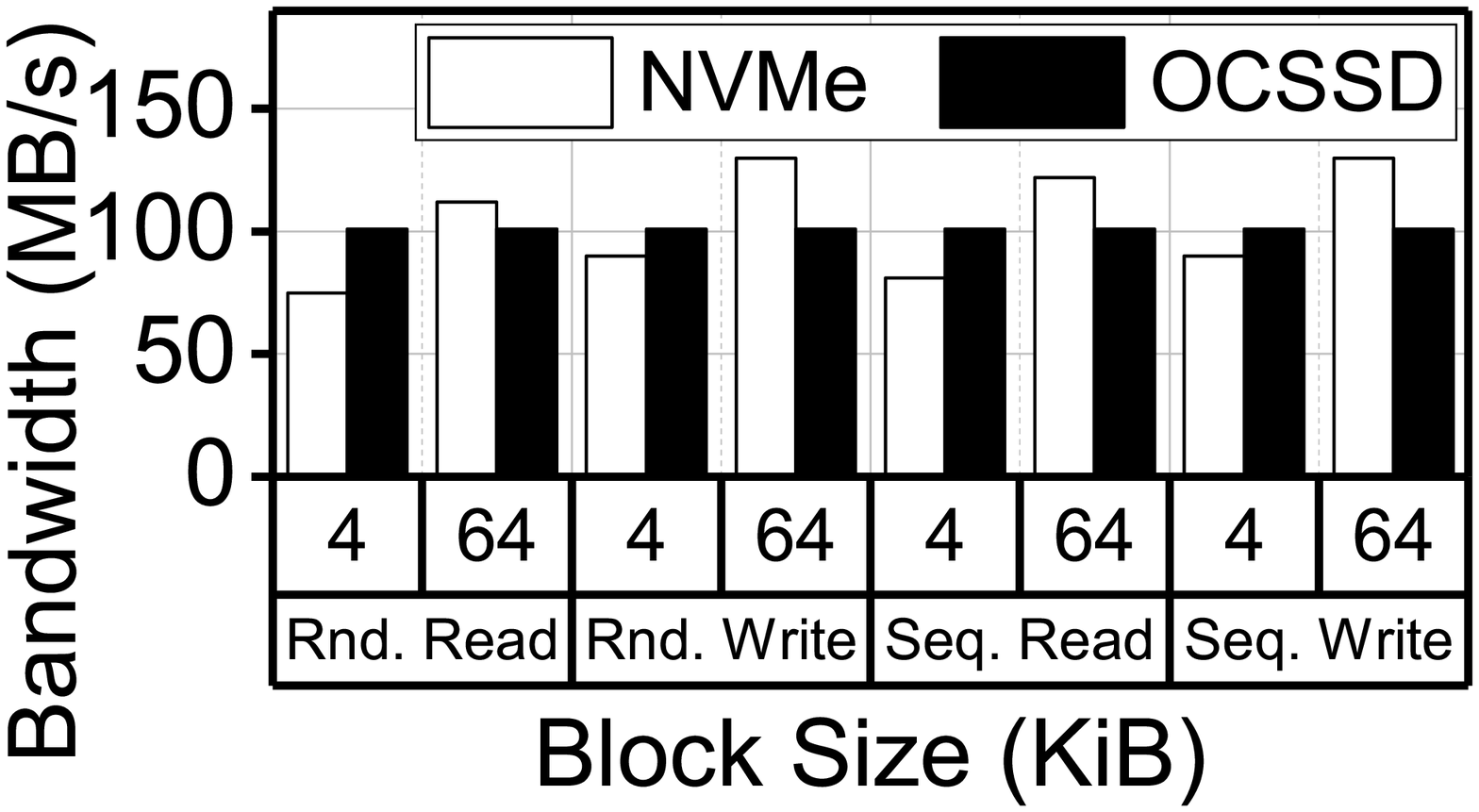}\hspace{5pt}}}
	\subfloat[Kernel CPU Utilization.]{\label{fig:ocssd_cpu}\rotatebox{0}{\includegraphics[width=0.33\linewidth]{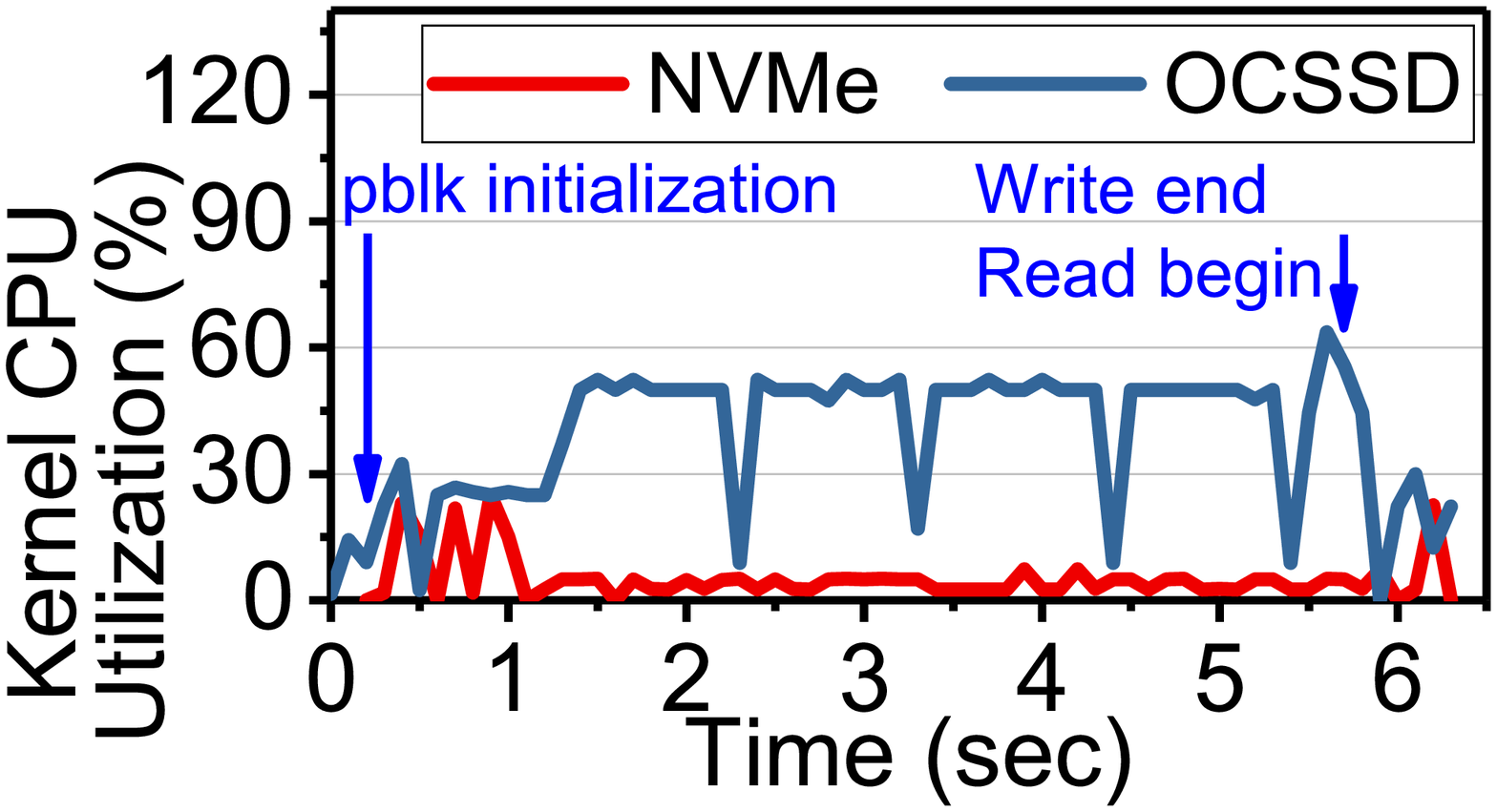}\hspace{5pt}}}
	\subfloat[Total DRAM usage.]{\label{fig:ocssd_dram}\rotatebox{0}{\includegraphics[width=0.33\linewidth]{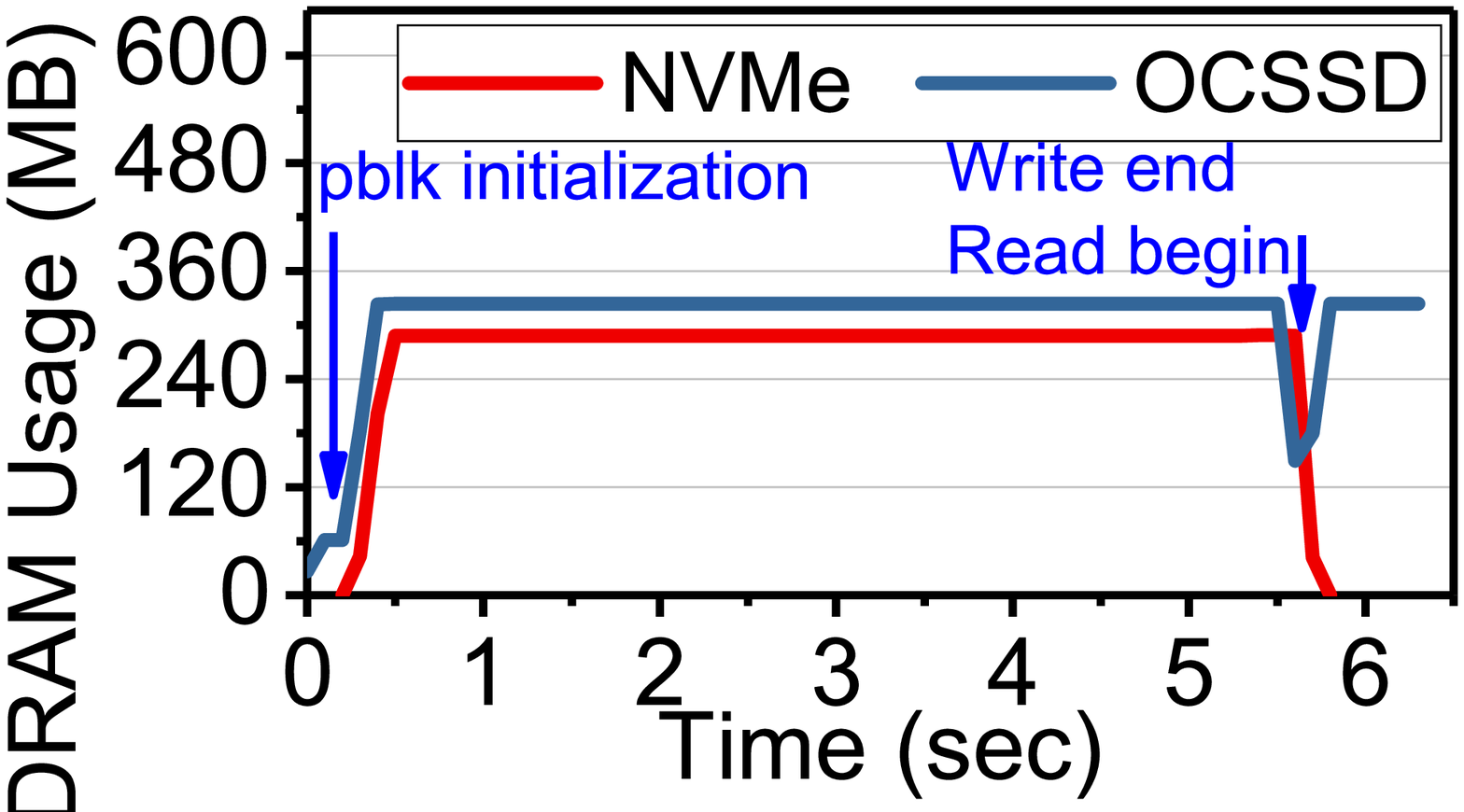}\hspace{5pt}}}
	\caption{Overall performance and system utilization of Active and Passive SSD.}
	\label{fig:ocssd}
	\end{minipage}
	\hspace{5pt}
\end{figure*}

\begin{table*}[ht]
	\begin{minipage}[b]{.79\linewidth}
	\centering
	\begin{adjustbox}{width=5.5in,center=\textwidth}
		\begin{threeparttable}
			\setlength\tabcolsep{5pt}
			\begin{tabular}{|c|cc|cccccc|cccc|cc|ccc|ccc|cc|ccc|cc|c|}
				\hline
				\rule{0pt}{10pt}
				\multirow{2}{*}{} & \multicolumn{2}{c|}{Mode} & \multicolumn{6}{c|}{H-CPU(ARM)} & \multicolumn{4}{c|}{Host interface} & \multicolumn{2}{c|}{C-Cplx\tnote{3}} & \multicolumn{3}{c|}{S-Cplx\tnote{4}} & \multicolumn{3}{c|}{Cache} & \multicolumn{2}{c|}{FTL} & \multicolumn{3}{c|}{Power} & \multicolumn{2}{c|}{Dynamics} & {Sup\tnote{13}}\\ \cline{2-29}
				\rule{0pt}{27pt}
				& \rotatebox[origin=c]{90}{SA\tnote{1}} & \rotatebox[origin=c]{90}{FS} & \rotatebox[origin=c]{90}{Atomic} & \rotatebox[origin=c]{90}{Timing} & \rotatebox[origin=c]{90}{Minor} & \rotatebox[origin=c]{90}{HPI\tnote{2}} & \rotatebox[origin=c]{90}{DerivO3} & \rotatebox[origin=c]{90}{O3\_v7a} & \rotatebox[origin=c]{90}{SATA} & \rotatebox[origin=c]{90}{UFS} & \rotatebox[origin=c]{90}{NVMe} & \rotatebox[origin=c]{90}{OCSSD} & \rotatebox[origin=c]{90}{CPU} & \rotatebox[origin=c]{90}{DRAM} & \rotatebox[origin=c]{90}{Tranx\tnote{5}} & \rotatebox[origin=c]{90}{SP/SB\tnote{6}} &
				\rotatebox[origin=c]{90}{ISPP\tnote{7}} &  \rotatebox[origin=c]{90}{Config\tnote{8}} & \rotatebox[origin=c]{90}{RA\tnote{9}} & \rotatebox[origin=c]{90}{Full\tnote{10}} & \rotatebox[origin=c]{90}{Hybrid} & \rotatebox[origin=c]{90}{Page\tnote{11}} & \rotatebox[origin=c]{90}{CPU} & \rotatebox[origin=c]{90}{DRAM} & \rotatebox[origin=c]{90}{NAND} & \rotatebox[origin=c]{90}{Energy} & \rotatebox[origin=c]{90}{Exec\tnote{12}} & \rotatebox[origin=c]{90}{Queue}\\ \hline
				Amber & \ding{51} & \ding{51} & \ding{51} & \ding{51} & \ding{51} & \ding{51} & \ding{51} & \ding{51} & \ding{51} & \ding{51} & \ding{51} & \ding{51} & \ding{51} & \ding{51} & \ding{51} & \ding{51} & \ding{51} & \ding{51} & \ding{51} & \ding{51} & \ding{51} & \ding{51} & \ding{51} & \ding{51} & \ding{51} & \ding{51} & \ding{51} & \ding{51}\\ \hline
				SimpleSSD 1.x\cite{simplessd}\cite{simplessd-x} & \ding{51} & \ding{51} & \ding{51} &  &  &  &  &  &  &  & \ding{51} &  &  & \ding{51} & \ding{51} & \ding{51} & \ding{51} & \ding{51} & \ding{51} & \ding{51} & \ding{51} &  &  &  & \ding{51} &  &  & \ding{51} \\ \hline
				MQSim\cite{mqsim} & \ding{51} &  &  &  &  &  &  &  & \ding{51} &  & \ding{51} &  &  & \ding{51} & \ding{51} &  & \ding{51} &  &  & \ding{51} & \ding{51} & \ding{51} &  &  &  &  &  & \ding{51} \\ \hline
				SSDSim\cite{ssdsim} & \ding{51} &  &  &  &  &  &  &  &  &  &  &  &  &  &  &  &  &  &  & \ding{51} & \ding{51} & \ding{51} &  &  &  &  &  &  \\ \hline
				SSD-Extension\cite{ssd-extension} & \ding{51} &  &  &  &  &  &  &  &  &  &  &  &  &  &  & \ding{51} &  &  &  &  &  & \ding{51} &  &  &  &  &  &  \\ \hline
				FlashSim\cite{flashsim} & \ding{51} &  &  &  &  &  &  &  &  &  &  &  &  &  &  & \ding{51} &  &  &  &  & \ding{51} & \ding{51} &  &  &  &  &  &  \\ \hline
			\end{tabular}
			\begin{tablenotes}[para]
				\item[1] Standalone\hspace{-7pt}
				\item[2] High perf. In-Order\hspace{-7pt}
				\item[3] Computation complex\hspace{-7pt}
				\item[4] Storage complex\hspace{-7pt}
				\item[5] Transaction scheduling\hspace{-7pt}
				\item[6] Super page/block\hspace{-7pt}
				\item[7] Incremental step pulse programming\hspace{-7pt}
				\item[8] Configurable cache\hspace{-7pt}
				\item[9] Readahead\hspace{-7pt}
				\item[10] Fully associative\hspace{-7pt}
				\item[11] Page-level mapping\hspace{-7pt}
				\item[12] Dynamic firmware execution\hspace{-7pt}
				\item[13] Support
			\end{tablenotes}
		\end{threeparttable}
	\end{adjustbox}
	\vspace{2pt}
	\caption{Feature comparison across various simulators.\label{tab:compare}}
	\end{minipage}
	\begin{minipage}[b]{.19\linewidth}
	\vspace{2pt}
	\centering
	\includegraphics[width=.90\linewidth]{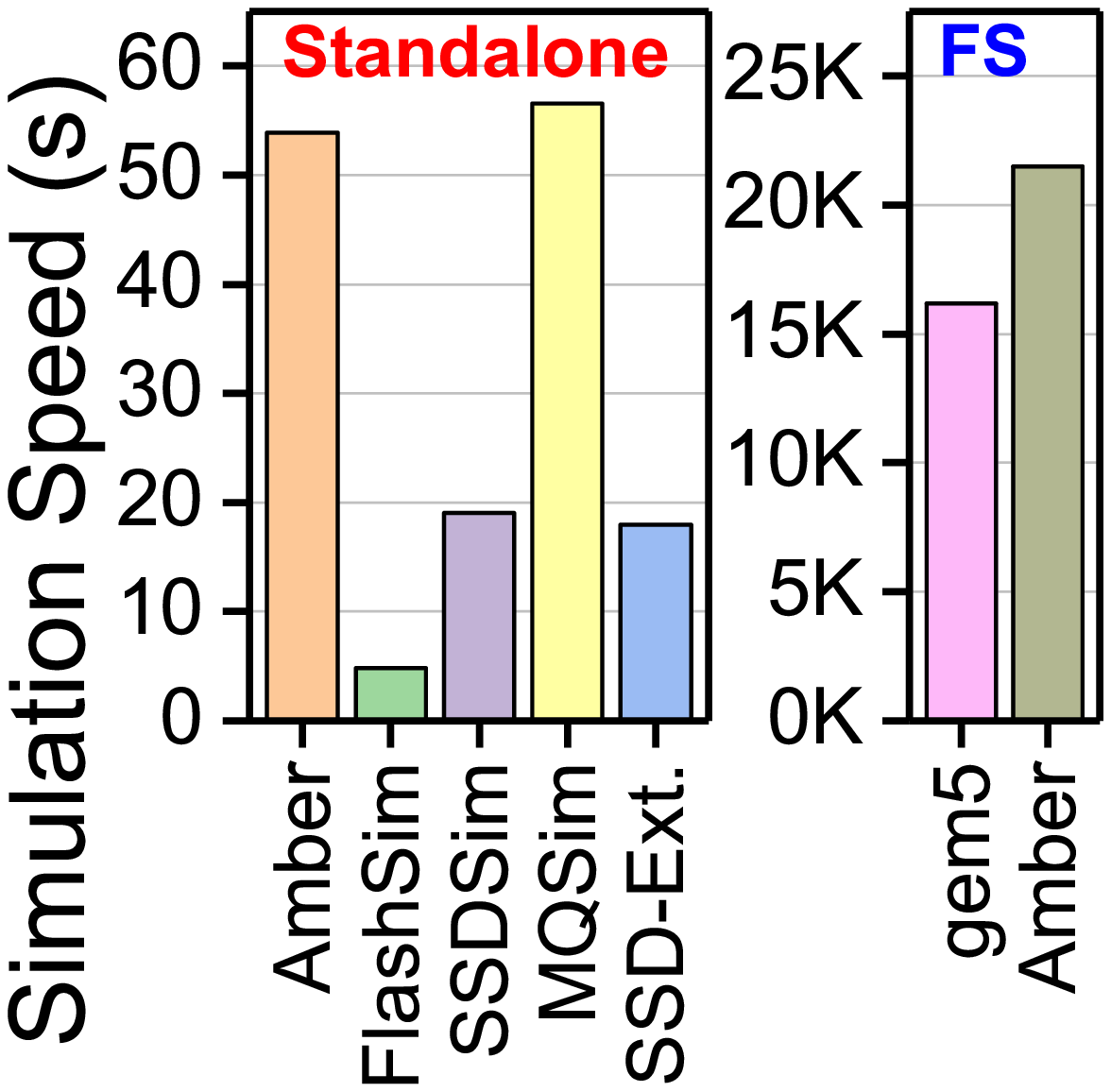}
	\captionof{figure}{Execution time.\label{fig:simtime}}
	\end{minipage}
	\vspace{-2pt}
\end{table*}

Figure \ref{fig:ocssd_bw} shows the performance of NVMe SSD (Active approach) and OCSSD 2.0 (Passive approach). As it can be observed, passive approach is better to service I/O requests than active approach in case of small I/O accesses (4KB-sized) with both random and sequential patterns. Specifically, OCSSD shows a 30\% better throughput than NVMe SSD for 4KB-sized requests. This is mainly because OCSSD is in a better position to utilize host-side buffer cache with better information, which can in turn directly serve the request from the host rather than going through the underlying storage. However, in case of large-sized I/O accesses (64KB), NVMe exhibits a 20\% better average throughput, due to the limited system buffer capacity that kernel-level drivers can use. Note that, in contrast to user-level memory, kernel memory is not freely used as it directly allocates the buffer from the physical memory address space instead of virtual memory space.

Figures \ref{fig:ocssd_cpu} and \ref{fig:ocssd_dram} plot the CPU utilization and memory requirement of OCSSD and NVMe devices, respectively. At the beginning of I/O processing, FIO consumes similar CPU cycles for both OCSSD and NVMe SSD due to write initialization for warming up processes. However, after the initialization, OCSSD consumes 50\% of all the cores (four) whereas NVMe SSD only uses 10\% of CPU. This is because, OCSSD requires to run \texttt{pblk} and \texttt{LightNVM} drivers to perform address translations, memory caching, and physical flash media management. On the other hand, \texttt{pblk} and \texttt{LightNVM} drivers do not use DRAM memory too much compared to NVMe SSD. As they are kernel drivers, OCSSD allocates system memory at the initialization phase, and reuses them for the entire execution phase (64 MB). Since FIO and NVMe protocol management basically requires about 280 MB of memory, the driver's memory requirement pressure can be ignored.


\section{Simulator Comparisons and Related Work}
\label{sec:relatedwork}

In literature, only a few simulators exist for high-performance SSD modeling. 
SSD Extension for DiskSim (SSD-Extension) \cite{ssd-extension} is the most popular simulator, which extends DiskSim \cite{disksim} by adding SSD models. This simulator simulates a page mapping FTL built upon a simplified flash model. FlashSim \cite{flashsim} tries to extend the mapping algorithm from the page-level to different associativities, but has no flash model or queue at its storage complex. In contrast to SSD-Extension, SSDSim \cite{ssdsim} can capture the details of internal parallelism by considering flash dies, planes and channel resources extracted by an in-house FPGA platform. However, SSDSim has no model for storage interface and the corresponding queue control mechanisms. MQSim \cite{mqsim} models a storage complex and enhances SSD simulation in comparison, by adding simple DRAM/flash and protocol management latency models, which partially consider the latency of the computation complexity. However, it cannot capture any internal embedded core latency, and has no capability of SSD emulation (that store/load actual data). As real contents and data movements are omitted in its model, MQSim cannot have a full storage stack on the host side; a system emulation mode of gem5, which only captures the system timing without having file system or storage stack might be possible to execute, but an SSD-enabled full system by having all software/hardware components for both host and storage cannot be explored. We checked all the repositories of their simulation framework and verified the unavailability of full-system simulations.  There are a few SSD simulations built on QEMU \cite{qemu}, which allow the SSD simulators communicate with a host emulation system. Unfortunately, FEMU \cite{femu} removes many necessary components of the storage stack to achieve a reasonable simulation speed, and VSSIM \cite{vssim} only supports IDE-based storage.
In contrast, SimpleSSD \cite{simplessd} can be attached to gem5 over IDE and emulate data transfers, and its extension \cite{simplessd-x} employs NVMe queue protocols that handle a pointer-based communication between host and storage-side simulations. However, these simulators cannot accommodate diverse storage interfaces and protocols atop different CPU timing and detailed memory models, which require all responses for software/hardware modules in a very-fine granular manner. In addition, none of the aforementioned simulators has the capability of capturing the SSD's computation complex, including dynamic firmware executions, and power/energy measurement.

Table \ref{tab:compare} summarizes the differences between the existing SSD simulators and our proposed \emph{Amber}. \emph{Amber} can capture embedded CPU cores' performance such as CPI and break down firmware executions into ``branch'', ``load/store'' and ``arithmetic'' operations, which none of the existing simulators is able to perform. In addition, \emph{Amber} can report dynamic power consumption of firmware stack executions, by taking into account the embedded cores, internal DRAMs, and flash devices. By modifying system barbus and implementing a DMA engine (that enables SSD emulation), \emph{Amber} can operate under both functional CPU and timing CPU models in full-system environment. It also models different host controllers (SATA/UFS) and drivers (NVMe/OCSSD), which can enable all s-type storage and h-type storage under diverse computing from mobile and general computing systems. Figure \ref{fig:simtime} compares simulation speeds of diverse standalone simulators (left), and pure gem5 and Amber (right). The simulation speed of Amber is slightly better than that of MQSim, and it captures all SSD-enabled full system characteristics with a reasonable simulation speed.



\section{Acknowledgement}
This research is mainly supported by NRF 2016R1C1B2015312, DOE DEAC02-05CH11231, IITP-2018-2017-0-01015, NRF 2015M3C4A7065645, Yonsei Future Research Grant (2017-22-0105) and MemRay grant (2015-11-1731). The authors thank Samsung’s Jaeheon Jeong,
Jongyoul Lee, Se-Jeong Jang and JooYoung Hwang for their SSD sample donations. N.S. Kim is supported in part by grants from NSF CNS-1557244 and  CNS-1705047. M. Kandemir is supported in part by grants by NSF grants 1822923,  1439021,
1629915, 1626251, 1629129, 1763681, 1526750 and 1439057. The simulator is designed, developed, and maintained by Computer Architecture and MEmory systems Laboratory (CAMELab). Myoungsoo Jung is the corresponding author.

\section{Conclusion}
\label{sec:conclusion}
We propose a new SSD simulation framework, Amber, which incorporates a full-featured SSD model into a full system environment and emulates all software stacks employed in diverse OS and hardware computing platforms. Amber modifies host buses, and implements plenty of storage interfaces and protocols such as SATA, UFS, NVMe, and OCSSD. 

\makeatletter
\def\endthebibliography{%
	\def\@noitemerr{\@latex@warning{Empty `thebibliography' environment}}%
	\endlist
}
\makeatother
\bibliographystyle{ieeetr}
\bibliography{ref}

\end{document}